# Surfing in the phase space of spin-orbit coupling in binary asteroid systems

Mahdi Jafari Nadoushan★
*K. N. Toosi University of Technology, Tehran, Iran*

**ABSTRACT**

For a satellite with an irregular shape, which is the common shape among asteroids, the well-known spin-orbit resonance problem could be changed to a spin-orbit coupling problem since a decoupled model does not accurately capture the dynamics of the system. In this paper, having provided a definition for close binary asteroid systems, we explore the structure of the phase space in a classical Hamiltonian model for spin-orbit coupling in a binary system. To map out the geography of resonances analytically and the cartography of resonances numerically, we reformulate a fourth-order gravitational potential function, in Poincare variables, via Stokes coefficients. For a binary system with a near-circular orbit, isolating the Hamiltonian near each resonance yields the pendulum model. Analysis of the results shows the geographical information, including the location and width of resonances, is modified due to the prominent role of the semi-major axis in the spin-orbit coupling model but not structurally altered. However, this resulted in modified Chirikov criterion to predict onset of large-scale chaos. For a binary system with arbitrary closed orbit, we thoroughly surf in the phase space via cartography of resonances created by fast Lyapunov indicator (FLI) maps. The numerical study confirms the analytical results, provides insight into the spin-orbit coupling, and shows some bifurcations in the secondary resonances which can occur due to material transfer. Also, we take the (65803) Didymos binary asteroid as a case to show analytical and numerical results.

## 1 INTRODUCTION

A spin-orbit resonance is usually defined as a situation where the rotation period of a satellite and its motion around its host body are commensurate. In this paper, when the spin of the satellite is studied while its orbit is assumed fixed, we call the problem spin-orbit resonance. When the simultaneous alternation of orbit and spin of the satellite is studied, we call it spin-orbit coupling. Goldreich & Peale (1966) introduced spin-orbit resonance in the solar system. Since, the spin-orbit resonance problem has been studied extensively. Wisdom et al. (1984) investigated the chaotic rotation of aspherical Saturn's satellite Hyperion using the spin-orbit resonance model through Chirikov's criterion for resonance overlapping. Celletti (1990) studied the stability of spin-orbit resonances using an approximated mathematical model and applying the Kolmogorov-Arnold-Moser theory. Celletti & Chierchia (2000) numerically studied the stability of periodic orbits of a nearly-integrable conservative spin-orbit resonance model through a symplectic algorithm that reduces the system to a map. Flynn & Saha (2005) studied the second-order spin-orbit resonances employing Lie transform perturbation theory. Jafari Nadoushan & Assadian (2015, 2016) thoroughly explored the phase space of spin-orbit resonance and examined the effects of system parameters on the chaos occurring in the spin-orbit resonance model. Misquero & Ortega (2020) investigated the spin-orbit problem analytically, considering dissipative forces without restriction on the orbital eccentricity.

Works mentioned above have considered the spin-orbit resonance model in which the orbit is fixed. Some other works have focused on a model with variable orbit. Breiter et al. (2005) studied spin-orbit coupling through the Kinoshita problem utilizing a Lie-Poisson integrator. Naidu & Margot (2015) examined the spin-orbit coupling of binary asteroids using the surface of section and showed the existence of a chaotic rotational motion. Hou & Xin (2017) considered fixed and variable orbits in the spin-orbit resonance. They showed that for the variable orbit, the resonance center could change for some values of the system parameters. Correia & Delisle (2019) presented a formulation of the dissipative spin-orbit coupling by considering gravitational tidal torques for close-in planet and studied the spin evolution. Wang & Hou (2020) considered a binary asteroid in the synchronous resonance with a variable orbit. They studied the spin of the secondary and its effect on the orbit. These works show that the variable orbit in spin evolution of the secondary is considered; nevertheless, the spin-orbit coupling model has not received considerable attention. Therefore, in this paper, we deal with the spin-orbit coupling in binary asteroid systems. In binary systems like binary asteroids, which have irregular shapes, the descriptive model needs to be upgraded from the spin-orbit resonance model to the spin-orbit coupling model. The irregular shape of the secondary has a significant effect on the mutual orbit. When the shape of the asteroid is not a sphere, the gravitational interaction will be important, and we should simultaneously consider the interaction of orbit and spin. Nevertheless, it seems that best criterion for utilizing either the coupling model or the decoupling model is the measurement of rotational and orbital angular momentum relative to each other, which leads to the definition of close binary asteroids.

In this paper, we focus on a binary asteroid with a primary, which we assume as a sphere, and a secondary as an ellipsoid. We calculate the potential energy function up to the fourth-order and based on sets of spherical harmonics (Stokes) coefficients. We employ the first fundamental resonances model for delineating the local structure of phase space near each resonance analytically. We provide a comprehensive analytical study for near-circular orbit through geographic information and numerical study for arbitrary closed orbit through cartographic images. In this way, we examine the phase space of spin-orbit coupling, which was not counted in the above tallies. To this end, two well-known analytical and numerical methods, resonance overlap criterion and Fast Lyapunov Indicator maps, are exploited. We present and find the following. (1) We introduce a criterion for the definition of *close binary asteroid systems* from the point of view of dynamical closeness of an asteroid pair. (2) We provide a comprehensive study of spin-orbit coupling for binary asteroids. (3) We find that ignoring the coupling of spin and orbit leads to wrong geographical information and phase portrait due to the role of semi-major axis. Particularly, due to the variable orbit and thus eccentricity, there may be a chaotic layer in the phase space. (4) Although the formulation of the spin-orbit coupling, unlike the spin-orbit resonance, says that the role of the equilibrium points might mathematically change through an alternation of the orbital semi-major axis, we found this physically impossible. (5) Considering spin-orbit coupling, the geographical information is modified and prediction of resonance overlapping is improved. We provide a modified criterion for onset of global chaos. (6) Expanding the gravitational potential function up to the fourth-order brings about new resonances which leads to improving prediction of resonance overlapping. (7) The 3:1 secondary resonance is abundant in binary asteroids and its effect can be studied on the dynamical behavior, especially tidal effects. However, some bifurcations of 3:1 secondary resonance are observed. The material transfer from one body to another and changing the shape of the secondary can lead to bifurcation.

★ E-mail: mjafari@kntu.ac.ir

This paper is organized as follows: in the section 2, the Hamiltonian of the spin-obit coupling is derived. The section 3 discusses a criterion for distinguishing between coupling and decoupling binary systems. The geography of the resonances is computed in the section 4. The cartography images of the phase space for different values of the parameters are presented in the section 5. In the section 6, the (65803) Didymos binary asteroid is investigated as a case. The section 7 gives the conclusion.

## 2 HAMILTONIAN OF THE SYSTEM

We provide the Hamiltonian of the spin-orbit coupling here for the planar case. Generally, the Hamiltonian of a natural system is the sum of kinetic and potential energy. The kinetic energy for the planar motion of an ellipsoid around a spherical rigid body is trivial. However, the potential energy is almost complicated. Many papers have derived the mutual gravitational potential energy function between two rigid bodies in different ways (Schutz (1981); Werner & Scheeres (2005); Ashenberg (2007); Compère & Lemaitre (2014)). We have expressed the potential energy function with MacCullagh's approach in Jafari Nadoushan & Assadian (2015, 2016). Here, we utilized the mutual gravitational potential energy function up to the fourth order in terms of Stokes coefficients. The higher order gravity harmonics such as the fourth-order are most important at small separations (Ćuk & Nesvornỳ (2010)). Considering the fourth-order gravity harmonic causes 4:3 and 4:5 resonances to appear in the analysis in addition to improving prediction of overlap of resonances.

We assume that the ellipsoidal asteroid of the mass $m_s$ is subjected to the gravitational attraction of a homogenous spherical asteroid with the mass $m_p$ and the center of mass of the ellipsoid moves on a variable orbit, as we show in Fig. 1. Therefore, the gravitational potential is:

$$V = -\int_{m_s} \frac{Gm_p}{\mathbf{R}} dm \tag{1}$$

where $G$ is the universal gravitational constant and $\mathbf{R} = \mathbf{r} + \boldsymbol{\rho}$, which $\mathbf{r}$ is the instantaneous orbital radius vector and $\boldsymbol{\rho}$ is the position vector of $dm$ with respect to the center of mass of the ellipsoid.

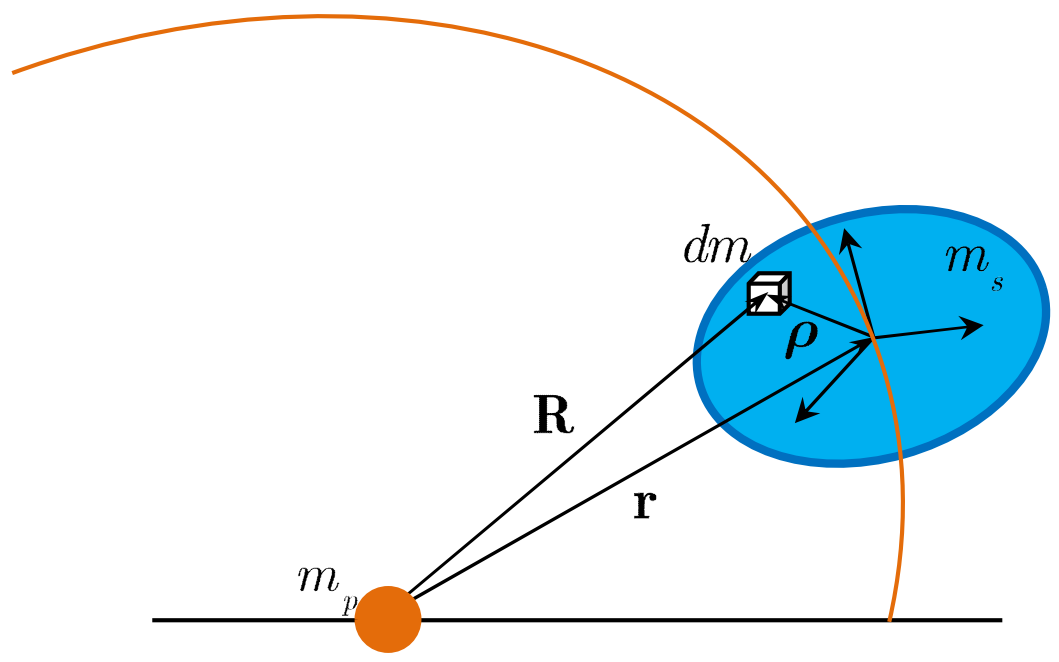

**Figure 1.** Geometry and notations of Sphere-ellipsoid system.

We could express the magnitude of $\mathbf{R}$ as

$$|\mathbf{R}| = R = r\sqrt{1 + \left(\frac{\rho}{r}\right)^2 + 2\frac{\mathbf{r}^T\boldsymbol{\rho}}{r^2}} \tag{2}$$

and, whereby, we expand the inverse of $R$ via Legendre polynomials, $P_l$ as follows:

$$\frac{1}{R} = \frac{1}{r\sqrt{1+\alpha^2 - 2\alpha q}} = \frac{1}{r}\sum_{l=0}^{\infty} (-\alpha)^l P_l\,[q] \tag{3}$$

where $\alpha = -\rho/r$ and $q = (\mathbf{r}^T\boldsymbol{\rho})/(r\rho)$. As a result, we rewrite the gravitational potential function as

$$V = -\frac{Gm_p}{r}\int_{m_s}\sum_{l=0}^{\infty}(-\alpha)^l P_l\,[q]\,dm \tag{4}$$

According to Tricarico (2008), we could explicitly calculate the above integral up to the fourth order, i.e., $l = 4$, for the planar sphere-ellipsoid system as follows:

$$V = -Gm_p \sum_{l=0}^{4} \frac{\widetilde{V}_l}{r^{l+1}} \tag{5}$$

Now, let us introduce the geometry of the spin-orbit coupling model. Fig. 2 shows the geometrical setup of a planar motion of an ellipsoidal asteroid around a sphere asteroid. We consider that the equatorial plane of the ellipsoidal asteroid, as the secondary, is coincident with the orbital plane. The moments of inertia of the ellipsoid about principal axes are $I_1 < I_2 < I_3$. In the spin-orbit coupling, unlike the spin-orbit resonance, the mutual orbit varies with variable semi-major axis $a$ and eccentricity $e$. We should note that in the spin-orbit resonance model, as periapsis is fixed, it is a reference for measuring the orbital and rotational angles of the ellipsoid. However, in the spin-orbit coupling model, the periapsis varies, and hereupon we need an appropriate reference for measuring them, as is depicted in Fig. 2.

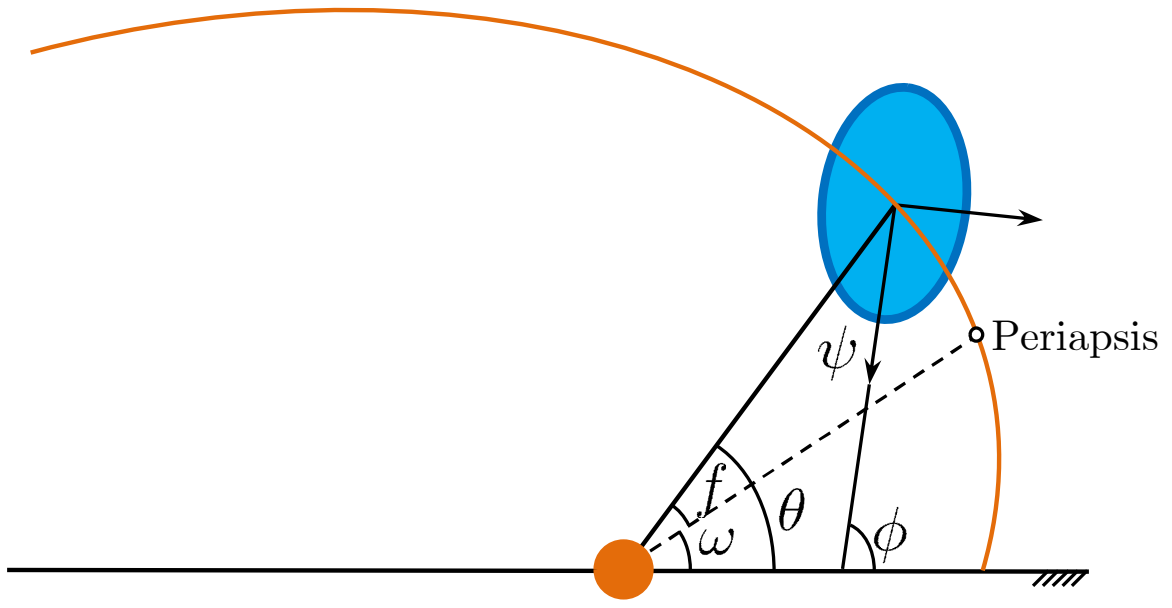

**Figure 2.** Geometry of spin-orbit coupling.

Now, we can write for $\widetilde{V}_l$ in equation (5):

$$
\begin{aligned}
\widetilde{V}_0 &= m_s \\
\widetilde{V}_2 &= -\frac{m_s a_s^2}{2} C_{20} + 3 m_s a_s^2 C_{22} \cos(2\psi) \\
\widetilde{V}_4 &= \frac{3 m_s a_s^4}{8} C_{40} + \frac{15 m_s a_s^4}{2} C_{42} \cos(2\psi) \\
&\quad + 105 m_s a_s^4 C_{44} \cos(4\psi)
\end{aligned}
\tag{6}
$$

where $\psi = \phi - \theta$ and $C_{ij}$ are Stokes coefficients and expressed as follows (Balmino (1994)):

$$
\begin{aligned}
C_{20} &= \frac{1}{5a_s^2}\left(c_s^2 - \frac{a_s^2+b_s^2}{2}\right) \\
C_{22} &= \frac{1}{20a_s^2}\left(a_s^2 - b_s^2\right) \\
C_{40} &= \frac{15}{7}\left(C_{20}^2 + 2C_{22}^2\right) \\
C_{42} &= \frac{5}{7} C_{20} C_{22} \\
C_{44} &= \frac{5}{28} C_{22}^2
\end{aligned}
\tag{7}
$$

where $a_s$, $b_s$, and $c_s$ are the semi-axes of the ellipsoid with $a_s \geq b_s \geq c_s$. Let us remark that the $\widetilde{V}_1$ is zero due to coinciding the origin with the center of mass, and the $\widetilde{V}_3$ vanishes owing to the symmetry of the ellipsoid. Finally, the gravitational potential energy function up to the fourth order is

$$
\begin{aligned}
V(r,\psi) = -G m_p m_s \Bigg[ & \frac{1}{r} - \frac{a_s^2 C_{20}}{2r^3} + \frac{3a_s^4 C_{40}}{8r^5} \\
& + \left(\frac{3a_s^2 C_{22}}{r^3} + \frac{15 a_s^4 C_{42}}{2r^5}\right)\cos(2\psi) \\
& + \frac{105 a_s^4 C_{44}}{r^5}\cos(4\psi)\Bigg]
\end{aligned}
\tag{8}
$$

Given Fig. 2, the rotational kinetic energy of the secondary that rotates about its maximum moment of inertia axis perpendicular to the orbital plane is $T_R = I_3\dot{\phi}^2/2$, and orbital kinetic energy is $T_o = m(\dot{r}^2 + r^2\dot{\theta}^2)/2$, where $m$ is $m_p m_s/(m_p + m_s)$ and is called reduced mass. Hence, the mass normalized Hamiltonian of the system is given below:

$$
\begin{aligned}
H(t, r, \dot{r}, \theta, \dot{\theta}, \phi, \dot{\phi}) = {} & \frac{1}{2}\left(\dot{r}^2 + r^2\dot{\theta}^2\right) - \frac{\mu}{r} + \frac{I_3}{2m}\dot{\phi}^2 \\
& -\mu\Bigg[-\frac{a_s^2 C_{20}}{2r^3} + \frac{3a_s^4 C_{40}}{8r^5} \\
& + \left(\frac{3a_s^2 C_{22}}{r^3} + \frac{15a_s^4 C_{42}}{2r^5}\right)\cos(2\phi - 2\theta) \\
& + \frac{105 a_s^4 C_{44}}{r^5}\cos(4\phi - 4\theta)\Bigg]
\end{aligned}
\tag{9}
$$

where $\mu = G(m_p + m_s)$ and $\theta$ measured from the reference is the sum of true anomaly $f$ and argument of perigee $\omega$. The first two terms describe the Keplerian motion, and the third term represents the free rotational motion. The last term is the asphericity perturbation, resulting in the orbital and rotational alteration of the secondary motion.

## 3 COUPLING OR DECOUPLING

The Hamiltonian (9) has no explicit time dependence so it is a conserved quantity. This means that the total energy of the system is conserved. Now, let us rewrite the Hamiltonian of the system in terms of generalized coordinates and momentums $(r, p_r, \theta, p_\theta, \phi, p_\phi)$ where $r$, $\theta$, and $\phi$ are the generalized coordinates, as are shown in Fig. 2 and $p_r$, $p_\theta$, and $p_\phi$ are the conjugate generalized momentums. According to the previous section, we can write

$$
H(r, p_r, \theta, p_\theta, \phi, p_\phi) = \frac{p_r^2}{2} + \frac{p_\theta^2}{2r^2} + \frac{p_\phi^2}{2mI_3} + V(r, \phi - \theta) \tag{10}
$$

We use an $F_2$-type generating function $F_2(\theta, \phi, p_\psi, p_\vartheta) = (\phi - \theta)p_\psi + \phi p_\vartheta$, which gives the following canonical transformation

$$
\begin{aligned}
\psi &= \phi - \theta \\
\vartheta &= \phi \\
p_\theta &= -p_\psi \\
p_\phi &= p_\psi + p_\vartheta
\end{aligned}
\tag{11}
$$

for canonical transfer of Hamiltonian (10) to

$$
\begin{aligned}
H(t, r, p_r, \psi, p_\psi, \vartheta, p_\vartheta) = {} & \frac{p_r^2}{2} + \left(\frac{1}{2r^2} + \frac{1}{2mI_3}\right)p_\psi^2 \\
& + \frac{p_\vartheta^2}{2mI_3} + \frac{p_\psi p_\vartheta}{mI_3} + V(r, \psi)
\end{aligned}
\tag{12}
$$

Because the new generalized coordinate $\vartheta$ does not appear in the Hamiltonian, it is an ignorable coordinate, and therefore, its conjugate generalized momentum $p_\vartheta$ is conserved. It should be noted that $p_\vartheta$ corresponds to the total angular momentum of the system, including orbital and rotational angular momentums i.e., $p_\vartheta = p_\theta + p_\phi$. Because there is no external torque, it is trivial that the total angular momentum is conserved. It is obvious that the orbital and rotational angular momentum can exchange with each other.

Let us look at the orbital and rotational angular momentums in comparison to each other. This comparison gives us a suitable criterion for identifying a condition in which the amount of angular momentum exchanged between the orbital and the rotational motion is negligible. When the secondary asteroid has negligible angular momentum compared to the mutual orbit, we can treat the spin and orbit to be decoupled. Nevertheless, if the orbital and rotational angular momentums are of the same order, the coupling becomes critical, and the orbital and spin evolutions are coupled. For example, chaotic rotation can induce chaos in orbital motion through the total angular momentum conservation.

For simplicity and without loss of generality, we assume that the secondary asteroid is a sphere. Thus, the ratio of rotational angular momentum to orbital angular momentum is

$$\frac{p_\phi}{p_\theta} = \frac{2}{5}\left(\frac{m_s}{m}\right)\left(\frac{a_s}{r}\right)^2\left(\frac{\dot\phi}{\dot\theta}\right) \tag{13}$$

We should note that the ratio of secondary mass to reduced mass can change between 1, corresponding to massless secondary, and 2, corresponding to a binary system with equal-mass components. The ratio of the radius of the secondary asteroid to orbital radius can vary from zero to one half. We consider the ratio of rotational angular velocity to orbital angular velocity between 0 and 5. Fig. 3 shows the ratio of rotational angular momentum to orbital angular momentum for various mass, size, and angular velocity ratios.

As seen from the figure, a secondary asteroid with a slow rotation and a large orbital radius has decoupled spin and orbit. The ratio of rotational angular momentum to orbital angular momentum increases by increasing the ratio of secondary mass to reduced mass. We can also see from the figure that when the orbital radius is ten times of radius of the secondary asteroid, the spin and orbit are decoupled for any ratio of secondary mass to reduced mass and even for fast rotation asteroid. If asteroids have separations within ten secondary radii, the rotational angular momentum is below ten percent of orbital angular momentum.

Binary asteroids can be in synchronous or double synchronous state (Pravec et al. (2016)) where the rotational angular velocity equals the orbital angular velocity. Fig. 4 shows the ratio of rotational angular momentum to orbital angular momentum for various ratios of secondary mass to reduced mass and size of secondary to orbital radius for synchronous binary asteroids. It is evident that as long as a binary system has inter-component separations within ten secondary radii, the momentum ratio is less than two percent for any ratio of secondary mass to reduced mass. Fig. 4 also depicts the ratios for eight binary systems having black triangle symbols. They include (175706) 1996 $FG_3$, (65803) Didymos, (66391) Moshup, (3749) Balam, (6369) 1983 UC, (8306) Shoko, (25021) Nischaykumar, (26416) 1999 XM84, (43008) 1999 UD31, (44620) 1999 RS43, (80218) 1999 VO123 (Pravec et al. (2019)). They have estimated inter-component separations from 5.5 to 9.7 times of secondary radii, and the estimated mass ratio of the secondary to the primary are between 0.015 to 0.097.

From Fig. 3 and Fig. 4, we conclude that we can offer a criterion for the definition of *close binary asteroid systems*, which actually is a dynamical closeness of asteroid pair. If the ratio of rotational angular velocity to orbital angular velocity is greater than two, and the mutual orbit of a binary asteroid system has a radius less than ten secondary radii, then we can categorize it in the close binary asteroid systems due to the comparable value of rotational angular momentum which results in spin and orbit coupling. For synchronous systems, if the orbital radius to the secondary radii is less than five, we can call the systems as a close one. In other words, the coupling of spin and orbit should lie at the heart of the definition of the closeness of binary asteroid systems. Accordingly, if the ratio of rotational angular momentum to the orbital angular momentum of a binary system is greater than ten percent, it is a close binary system. Therefore, by definition, all eight binary systems mentioned above are not close binary systems.

## 4 GEOGRAPHY OF RESONANCES

In this section, we analytically compute the location and the width of resonances as geographical information of resonances existing in the spin-orbit coupling model. To explore the geography of the resonances, we describe the local structure of resonances by the generic pendulum model or first fundamental resonance model. We usually average out the short-period terms from the Hamiltonian to investigate the resonant dynamics, which results in the averaged resonant Hamiltonian.

We define the unit of mass such that $m = 1$, and take $a_s$ as the unit of length. We also choose the unit of time such that $\mu = 1$. Then we introduce planar Poincare action-angle variables, which have no singularity for near-circular orbit, for representing orbits. The planar Poincare variables are defined by

$$\begin{aligned} \lambda &= M + \omega \\ \Lambda &= \sqrt{\overline{a}} \\ \gamma &= -\omega \\ \Gamma &= \sqrt{\overline{a}}\left(1 - \sqrt{1-e^2}\right) \end{aligned} \tag{14}$$

where $\overline{a}$ is length normalized orbital semi-major axis, and $M$ is the mean anomaly of the secondary. Also, we use generalized coordinate and momentum for representing rotational motion. Note that we hereafter drop the superscript for all normalized variables and parameters for notational simplicity.

We use a mathematical manipulation (see Jafari Nadoushan & Assadian (2016) for details) to derive the Hamiltonian of spin-orbit coupling in action-angle variables. Let us consider the critical argument $\chi_{k_s:k_o} = k_s\phi - k_o\lambda - (k_o - k_s)\gamma$ as the resonant angle for the $k_s : k_o$ spin-orbit resonance. If we consider up to infinite order of both gravitational potential energy function and elliptic expansions (Murray & Dermott (1999)), the compact form of multi-harmonic Hamiltonian function can be written as follows:

$$H = -\frac{1}{2\Lambda^2} + \frac{\Phi^2}{2I_3} - \sum_{k_s,k_o\in\mathbb{Z}}\sum_{i=0}^{\infty} X_{i,k_s:k_o}(\Lambda,\Gamma)\cos\left(i\chi_{k_s:k_o}\right) \tag{15}$$

where $\Phi$ is the action of angle $\phi$, $i$ is the number of harmonics, and $X_{i,k_s:k_o}$ is the coefficient of $i^{\text{th}}$ harmonic of the critical argument for the $k_s : k_o$ spin-orbit resonance and is the functions of the action variables $\Lambda$ and $\Gamma$; indeed orbital semi-major axis and eccentricity.

We are interested in the Hamiltonian for near-circular orbit up to the first order of eccentricity and with gravitational potential energy function up to the fourth order. For near-circular orbit, we approximate the action $\Gamma$ as $\Gamma = \sqrt{a}\left(1-\sqrt{1-e^2}\right) \approx \Lambda e^2/2$. Therefore, the Hamiltonian is

$$\begin{aligned} H\left(t,\lambda,\Lambda,\gamma,\Gamma,\phi,p_\phi\right) = &-\tfrac{1}{2\Lambda^2} + \tfrac{\Phi^2}{2}I_3 + \tfrac{4C_{20}\Lambda^4-3C_{40}}{8\Lambda^{10}} \\ &+\tfrac{12C_{20}\Lambda^4-15C_{40}}{8\Lambda^{10}}\sqrt{\tfrac{2\Gamma}{\Lambda}}\cos(\lambda+\gamma) \\ &+\tfrac{6C_{22}\Lambda^4-15C_{42}}{4\Lambda^{10}}\sqrt{\tfrac{2\Gamma}{\Lambda}}\cos(2\phi-\lambda+\gamma) \\ &-\tfrac{6C_{22}\Lambda^4+15C_{42}}{2\Lambda^{10}}\cos(2\phi-2\lambda) - \tfrac{105C_{44}}{\Lambda^{10}}\cos(4\phi-4\lambda) \\ &-\tfrac{42C_{22}\Lambda^4+135C_{42}}{4\Lambda^{10}}\sqrt{\tfrac{2\Gamma}{\Lambda}}\cos(2\phi-3\lambda-\gamma) \\ &+\tfrac{315C_{44}}{2\Lambda^{10}}\sqrt{\tfrac{2\Gamma}{\Lambda}}\cos(4\phi-3\lambda+\gamma) \\ &-\tfrac{1365C_{44}}{2\Lambda^{10}}\sqrt{\tfrac{2\Gamma}{\Lambda}}\cos(4\phi-5\lambda-\gamma) \end{aligned} \tag{16}$$

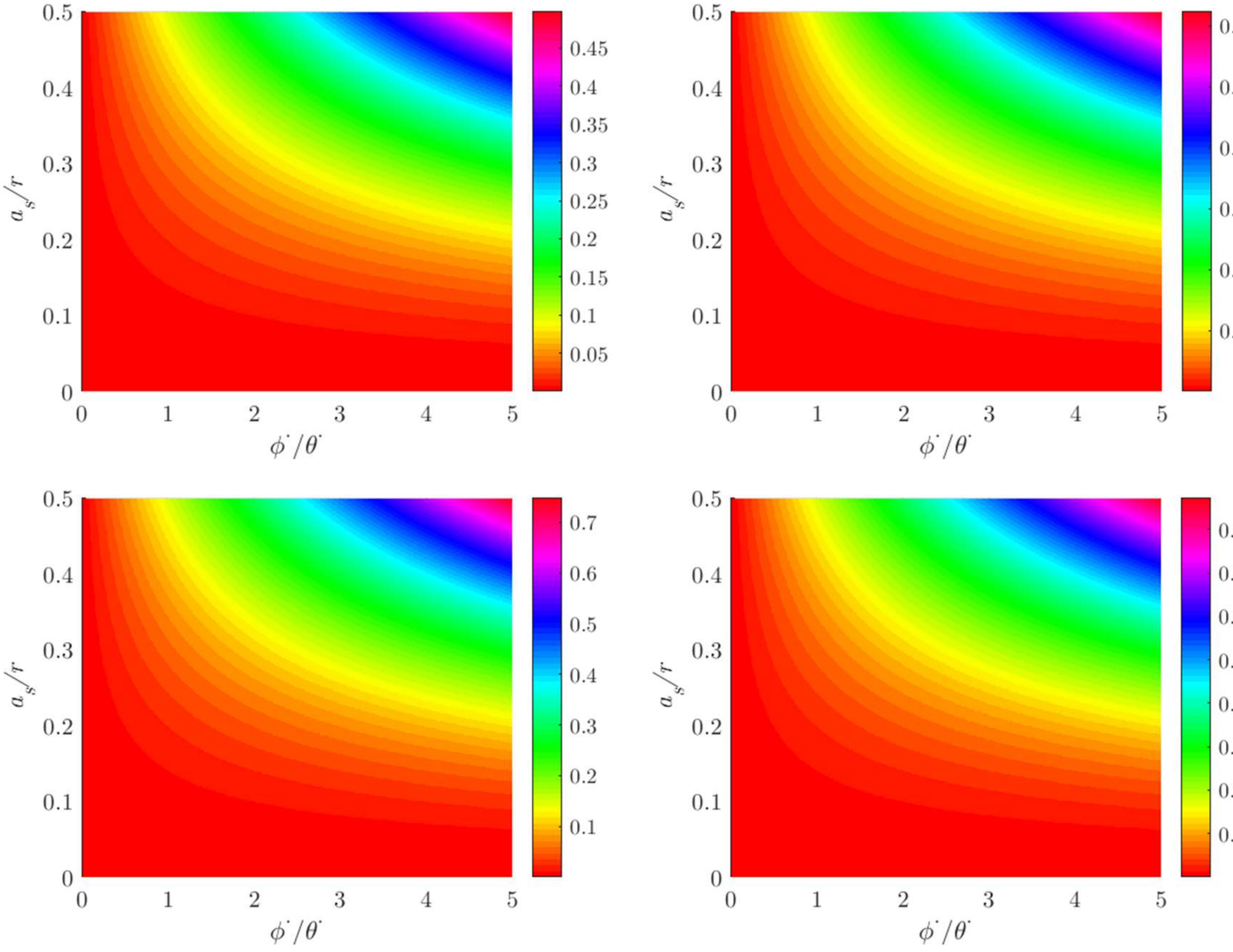

**Figure 3.** The ratio of rotational angular momentum to orbital angular momentum for various values of the ratios $m_s/m = 1$ (left top panel), $m_s/m = 1.25$ (right top panel), $m_s/m = 1.5$ (left bottom panel), $m_s/m = 1.75$ (right bottom panel). The color bar shows the ratio of rotational angular momentum to orbital angular momentum.

It is evident that, from equation (16), when we consider variable orbit, the existence of resonances is invariant. Because there are 1:1, 2:1, 2:3, 4:3, and 4:5 resonances in the Hamiltonian (16) as exist in the spin-orbit resonance model (Jafari Nadoushan & Assadian (2016)). Except 1:1, the others are first-order resonances as defines as $|k_s - k_o| = 1$. However, as will be seen, the geographical information of resonances could be modified. We remark that the appearance of resonances depends on the order of both orbital eccentricity and gravitational potential function. Moreover, considering higher-order terms of gravitational potential function results in a multi-harmonic Hamiltonian function.

We could write the Hamiltonian for the zero-order 1:1 or synchronous resonance as follows:

$$H_{1:1} = -\frac{1}{2\Lambda^2} + \frac{\Phi^2}{2I_3} - \sum_{i=0}^{2} \mathrm{X}_{i,1:1} \cos\left(2i\left(\phi - \lambda\right)\right) \tag{17}$$

and the Hamiltonian for the first-order resonances is

$$\begin{aligned} H_{k:k\pm1} = &-\tfrac{1}{2\Lambda^2} + \tfrac{\Phi^2}{2I_3} \\ &-\sum_{i=0}^{1} \mathrm{X}_{i,k:k\pm1} \cos\left(i\left(k\phi - (k\pm1)\lambda \mp \gamma\right)\right) \end{aligned} \tag{18}$$

here $k \in \{2, 4\}$. Note that near each resonance, other terms become nil through averaging over fast angles. Unlike the first-order resonances, the 1:1 resonance has two harmonics in the Hamiltonian. The synchronous resonance is the main resonance among the spin-orbit resonances and is evidently present even for circular orbits. This resonance is found in abundance in the solar system (Malhotra (1994)).

Let us continue with equation (18). We introduce a new set of canonical variables through the following generating function:

$$F_2 = k\phi\Psi_1 + \lambda\left(\Psi_3 - (k\pm1)\Psi_1\right) + \gamma\left(\Psi_2 \mp \Psi_1\right) \tag{19}$$

gives the following canonical transformation

$$\begin{aligned} &\Psi_1 = \tfrac{1}{k}\Phi && \psi_1 = k\phi - (k\pm1)\lambda \mp \gamma \\ &\Psi_2 = \Gamma \pm \tfrac{1}{k}\Phi && \psi_2 = \gamma \\ &\Psi_3 = \Lambda + \tfrac{k\pm1}{k}\Phi && \psi_3 = \lambda \end{aligned} \tag{20}$$

where $\Psi_i$ and $\psi_i$ are new conjugated action-angle variables. In terms of the new set of variables, the resonant Hamiltonian is expressed by

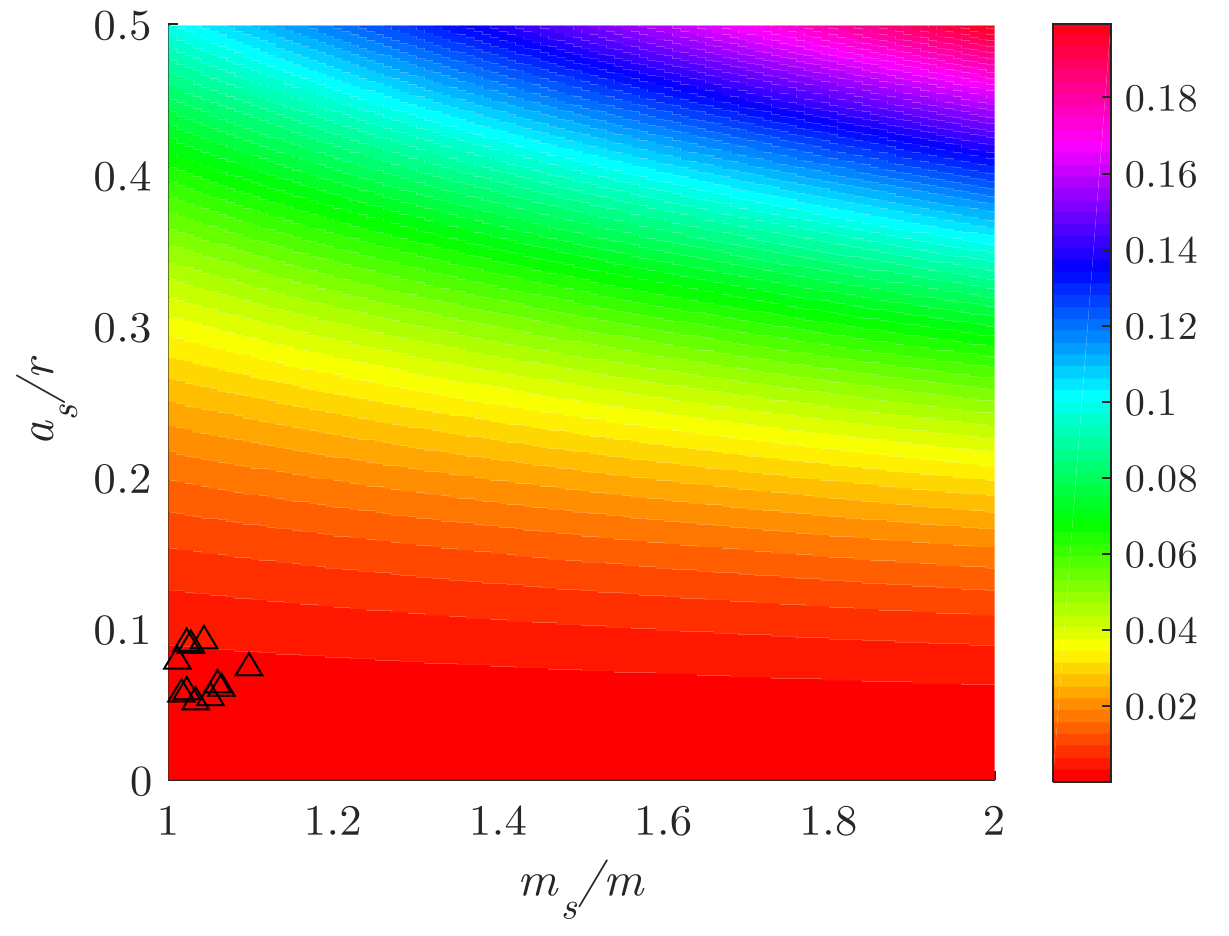

**Figure 4.** The ratio of rotational angular momentum to orbital angular momentum for synchronous binary asteroid and various values of size and ratio of secondary mass to reduced mass

.

$$H_{k:k\pm1} = -\frac{1}{2(\Psi_3-(k\pm1)\Psi_1)^2} + \frac{(k\Psi_1)^2}{2I_3} - \sum_{i=0}^{1} \mathrm{X}_{i,k:k\pm1}(\Psi_1,\Psi_2,\Psi_3)\cos(\mathrm{i}\psi_1) \tag{21}$$

Evidently, angular variables $\psi_2$ and $\psi_3$ are cyclic coordinates. Therefore, the actions $\Psi_2$ and $\Psi_3$ are all constants and play the role of parameters. $\Psi_2 = \sqrt{a}\left(1-\sqrt{1-e^2}\right) \pm \frac{1}{k} I_3\dot{\phi}$, apart from the near-circular approximation, is the total angular momentum. It means that the rotational angular momentum can exchange with that of the orbit through the alternation of orbit and rotation of the secondary in the long-term evolution. The exchange could be in such a way that the spin-orbit resonance may remain stable. This is called spin-orbit coupling, in which there is coupling between the orbital semi-major axis, eccentricity, and rotation of the secondary.

For synchronous resonances, we utilize a different generating function as follows:

$$F_2 = \phi\Psi_1 + \lambda(\Psi_3 - \Psi_1) \tag{22}$$

and thus, we obtain the resonant Hamiltonian as

$$H_{1:1} = -\frac{1}{2(\Psi_3-\Psi_1)^2} + \frac{\Psi_1^2}{2I_3} - \sum_{i=0}^{2} \mathrm{X}_{i,1:1}(\Psi_1,\Psi_3)\cos(2\mathrm{i}\psi_1) \tag{23}$$

As far as one can tell, the action $\Gamma$ is absent for synchronous resonance. Therefore, the orbital eccentricity does not have any contribution to the motion near zero-order resonance. The constant action $\Psi_3 = \Phi + \Lambda$ tells us the rotational angular momentum of the secondary and its Keplerian energy could exchange with each other so that, the energy of its orbit decreases with increasing rotation of the secondary. This leads to the orbit shrinking and consequently increases the orbital mean motion. This may be preserving the ratio of rotational velocity and orbital mean motion. Then, we will see a spin-orbit coupling in which the system stays in the exact synchronous resonance. It is worth mentioning that during the exchange, the orbit can be eccentric, and therefore the secondary liberates around the synchronous resonance.

However, remaining the system in the exact resonance seems unlikely, and the angular velocity and orbital semi-major axis of the secondary deviate from nominal commensurability. Hereupon, there is an excursion in action around the exact value of the action in resonance. The excursion may be approximated via expanding the Hamiltonian in an order three Taylor series about the exact resonance (Morbidelli (2002)):

$$\Delta H_{1:1} = \frac{1}{2}\left(\frac{1}{I_3} - \frac{3}{\left(\Psi_3-\Psi_1^{\mathrm{res}}\right)^4}\right)\delta\Psi_1^2 - \sum_{i=1}^{2} \mathrm{X}_{i,1:1}\left(\Psi_1^{\mathrm{res}},\Psi_3\right)\cos(2\mathrm{i}\psi_1) \tag{24}$$

in which the constant terms are removed from the Hamiltonian. The resulted Hamiltonian is very similar to that of a simple pendulum. Referring to Nadoushan & Assadian (2016), the second harmonic has no contribution to the width of the resonance. Consequently, the approximated amplitude or half-width of the synchronous resonant island in the spin-orbit coupling obtain as

$$\delta\Psi_1^{1:1} = \sqrt{\frac{12I_3C_{22}\left(\Psi_3-\Psi_1^{\mathrm{res}}\right)^4+30I_3C_{42}}{\left(\Psi_3-\Psi_1^{\mathrm{res}}\right)^{10}-3I_3\left(\Psi_3-\Psi_1^{\mathrm{res}}\right)^6}} \tag{25}$$

Similar to synchronous resonance we expand the Hamiltonian for first-order resonance, as follows:

$$\Delta H_{k:k\pm1} = \frac{1}{2}\left(\frac{k}{I_3} - \frac{3(k\pm1)^2}{\left(\Psi_3-(k\pm1)\Psi_1^{\mathrm{res}}\right)^4}\right)\delta\Psi_1^2 - \mathrm{X}_{1,k:k\pm1}\left(\Psi_1^{\mathrm{res}},\Psi_2,\Psi_3\right)\cos(\psi_1) \tag{26}$$

and the half-width of $k:k\pm1$ resonant island is approximated as

$$\delta\Psi_1^{k:k\pm1} = 2\sqrt{\frac{\mathrm{X}_{1,k:k\pm1}\left(\Psi_1^{\mathrm{res}},\Psi_2,\Psi_3\right)}{\frac{k}{I_3}-\frac{3(k\pm1)^2}{\left(\Psi_3-(k\pm1)\Psi_1^{\mathrm{res}}\right)^4}}} \tag{27}$$

Compared with the spin-orbit resonance model, the distinction of the half-width of resonances is due to the second term in the parenthesis in equation 24 and equation 26. This term appears because the spin-orbit coupling model includes variations in orbit. Therefore, in the spin-orbit coupling model, the width of resonances is modified.

The half-width of the resonances in terms of normalized orbital semi-major axis and eccentricity and physical parameters of the system are formulated in Table 1 for both spin-orbit coupling and spin-orbit resonance models. Manifestly, the orbital semi-major axis has a more prominent role in the spin-orbit coupling model than in the spin-orbit resonance model. The amplitude of resonances is modified and seems to have increased. For example, in a binary asteroid with 0.2 as the mass ratio, the normalized semi-major axis 3, the orbital eccentricity 0.05, and the ellipsoidal secondary with 1:0.9:0.8 as the normalized semi-axes, the amplitude of synchronous resonance in the spin-orbit coupling model increases about 8.1% compared to that of the spin-orbit resonance model. This value reaches about 11.7% for the 2:3 resonance. Now that the width of resonances has changed, the overlap of resonances might happen with moderate orbital elements in the spin-orbit coupling. The more irregular the second asteroid, i.e., the larger $C_{22}$, $C_{42}$, and $C_{44}$, the wider the resonances. The closer the mass of the primary asteroid and the mass of the secondary, the larger the width of resonances. Also, for the first-order resonances, a more eccentric orbit intensifies the strength of the resonances. Otherwise, the phase space is dominated by a synchronous resonant island.

After approximating the width of resonances, it is time to compute the equilibrium points and center of resonance regions. First of all, let us mention that Since the secondary is symmetric and the action of the symmetry group SO(2) leaves the system invariant, the description of orientation can be confined to $[0, \pi]$ interval. The equilibrium points of spin-orbit coupling are the same as for spin-orbit resonance mentioned in Jafari Nadoushan & Assadian (2016). For the 2:1, 1:1, and 2:3 resonances, the equilibrium points are at $\phi = 0, \pi/2, \pi$ at periapsis of the orbit. For the 4:3 and 4:5 resonances, $\phi = 0, \pi/4, \pi/2, 3\pi/4, \pi$ are the equilibrium points at the periapsis of the orbit. Some of these points are centers that are stable equilibrium points, and some are saddles that are unstable equilibrium points. Note that the resonance island has an almond shape with the maximum width at the center equilibrium point as the center of the resonance region.

**Table 1.** Half-width of the resonances for both spin-orbit coupling and spin-orbit resonance models.

| Resonance | Spin-Orbit Coupling | Spin-Orbit Resonance |
|---|---|---|
| 2:1 | $\sqrt{\frac{(6C_{22}a^2-15C_{42})e}{\frac{10a^5}{\left(1+\frac{m_s}{m_p}\right)\left(1+\left(\frac{b_s}{a_s}\right)^2\right)}-3a^3}}$ | $\sqrt{\frac{(6C_{22}a^2-15C_{42})e}{\frac{10a^5}{\left(1+\frac{m_s}{m_p}\right)\left(1+\left(\frac{b_s}{a_s}\right)^2\right)}}}$ |
| 1:1 | $\sqrt{\frac{12C_{22}a^2+30C_{42}}{\frac{5a^5}{\left(1+\frac{m_s}{m_p}\right)\left(1+\left(\frac{b_s}{a_s}\right)^2\right)}-3a^3}}$ | $\sqrt{\frac{12C_{22}a^2+30C_{42}}{\frac{5a^5}{\left(1+\frac{m_s}{m_p}\right)\left(1+\left(\frac{b_s}{a_s}\right)^2\right)}}}$ |
| 2:3 | $\sqrt{\frac{(42C_{22}a^2+135C_{42})e}{\frac{10a^5}{\left(1+\frac{m_s}{m_p}\right)\left(1+\left(\frac{b_s}{a_s}\right)^2\right)}-27a^3}}$ | $\sqrt{\frac{(42C_{22}a^2+135C_{42})e}{\frac{10a^5}{\left(1+\frac{m_s}{m_p}\right)\left(1+\left(\frac{b_s}{a_s}\right)^2\right)}}}$ |
| 4:3 | $\sqrt{\frac{630C_{44}e}{\frac{20a^5}{\left(1+\frac{m_s}{m_p}\right)\left(1+\left(\frac{b_s}{a_s}\right)^2\right)}-27a^3}}$ | $\sqrt{\frac{630C_{44}e}{\frac{20a^5}{\left(1+\frac{m_s}{m_p}\right)\left(1+\left(\frac{b_s}{a_s}\right)^2\right)}}}$ |
| 4:5 | $\sqrt{\frac{2730C_{44}e}{\frac{20a^5}{\left(1+\frac{m_s}{m_p}\right)\left(1+\left(\frac{b_s}{a_s}\right)^2\right)}-75a^3}}$ | $\sqrt{\frac{2730C_{44}e}{\frac{20a^5}{\left(1+\frac{m_s}{m_p}\right)\left(1+\left(\frac{b_s}{a_s}\right)^2\right)}}}$ |

Considering the first two terms in equation 17 and equation 18, it seems that in the spin-orbit coupling, unlike the spin-orbit resonance, the role of the equilibrium points might change through an alternation of the orbital semi-major axis. Indeed, the sign of the sum of these two terms determines the location of center of the resonance. When the spin of the secondary and its orbit are considered, the locations of center of resonances in the spin-orbit coupling are similar to that of the spin-orbit resonance unless the value of the normalized semi-major axis is less than two. Thus, it is impossible. However, this issue can be checked for the coupling between the spin of the primary and the orbit.

It must be noted that the isolated resonance hypothesis is true if physical and orbital parameters are in such a way that the sum of the half-width of adjacent resonances is less than their separation. Otherwise, KAM curves are destructed, and the resonance islands overlap, resulting in widespread chaos in the phase space where averaging method is not applicable. In this case, Chirikov resonance-overlap criterion (Chirikov (1979)) could be used to predict values of physical and orbital parameters for which onset of large-scale chaos occurs. According to separation of 1:1 and 2:3 resonances, and considering width of 4:5 resonance, we can estimate the onset of chaos by

$$\sqrt{\frac{12C_{22}a^2+30C_{42}}{\frac{5a^5}{\left(1+\frac{m_s}{m_p}\right)\left(1+\left(\frac{b_s}{a_s}\right)^2\right)}-3a^3}} + 2\sqrt{\frac{2730C_{44}e}{\frac{20a^5}{\left(1+\frac{m_s}{m_p}\right)\left(1+\left(\frac{b_s}{a_s}\right)^2\right)}-75a^3}} + \sqrt{\frac{(42C_{22}a^2+135C_{42})e}{\frac{10a^5}{\left(1+\frac{m_s}{m_p}\right)\left(1+\left(\frac{b_s}{a_s}\right)^2\right)}-27a^3}} = \frac{1}{2} \quad (28)$$

As it is evident, in the spin-orbit coupling the criterion depends on mass ratio, semi-axis ratios, orbital semi-major axis, and eccentricity. It is obvious that this criterion is more intricate than what Wisdom et al. (1984) have suggested.

## 5 CARTOGRAPHY OF RESONANCES

In the previous section, we limited ourselves to the near-circular orbit and assumed that the resonances are isolated and do not have any interaction with each other. In general, the coexistence of regular and chaotic motion in a system such as the system studied in this paper prevents a fully analytical description, and numerical exploration is necessary. In this section, we stroll through the phase space of the spin-orbit coupling through cartography of resonances crafted via FLI maps. The FLI map is a straightforward way to find a general image of the phase space of the problem.

The FLI, introduced by Froeschlé et al. (1997), is usually employed to discriminate between chaotic and regular motions and detect phase-space structure. For regular motion, the FLI grows linearly, while for a chaotic motion, the FLI increases exponentially. Larger FLI values indicate stronger chaos. Accordingly, we could distinguish between chaotic and regular motions.

Using Hamiltonian (9), which is non-dimensionalized and written in terms of the canonical variables, we could derive the equations of motion by applying the canonical formulation.

$$\dot{\mathbf{x}} = [\mathbf{x}, H] = \mathbf{f}(\mathbf{x}) \quad (29)$$

where $\mathbf{x}$ is a vector including the generalized coordinates and the conjugated generalized momentums, and $[,]$ is the Poisson bracket. The variational equations associated with the equations of motion are defined as

$$\dot{\mathbf{v}} = D\mathbf{f}(\mathbf{x})\,\mathbf{v} \quad (30)$$

where $D$ is the Jacobian operator. We use the equations of motion with associated variational equations to calculate the value of the FLI at time $t$ as follows (Skokos & Manos (2016)):

$$\mathrm{FLI}_t\left(\mathbf{x}(0), \mathbf{v}(0)\right) = \log\frac{\|\mathbf{v}(t)\|}{\|\mathbf{v}(0)\|} \quad (31)$$

We describe numerically the structure of the phase space of the spin-orbit coupling via FLI maps. We consider a set of 300×300 initial conditions located on a regular grid in the plane of normalized $\dot{\phi}$ and $\phi$, where normalized angular velocity ranges from 0 to 2.5, while the angle $\phi$ ranges from 0 to $\pi$. The integration time is a crucial issue. We find that the appropriate integration time is 100 times the non-dimensional orbital period. We report the FLI via a color code in the plane of initial conditions, such that the highest FLI values are shown in brown.

Before merely exploring the phase space of the spin-orbit coupling model, let us take a comparative look at the phase space of both spin-orbit coupling and spin-orbit resonance models for the same binary system. In Fig. 5, we show the phase space portrait of the spin-orbit coupling model in the left panel and the spin-orbit resonance model in the right panel for a system with values of $a = 3$, $e = 0.05$, $\theta = 0$, $m_s/m_p = 0.2$, and $(a_s : b_s : c_s) = (1 : 0.9 : 0.8)$. With the color-scale palette, as said before, the blue color characterizes a low FLI and indicates regular dynamics, whereas the brown color denotes a high FLI and indicates chaotic dynamics. It is clear that the phase portrait shown in Fig. 5 is much more complex than that of the pendulum model for the isolated resonance assumption.

Though the phase spaces are for the same system, it is evident that they are different. As we have seen from the analytical result and the numerical result confirms that the synchronous resonant island in the spin-orbit coupling model is wider than that of the spin-orbit resonance model. Although the width of the 2:3 resonant island in the spin-orbit resonance model is wider, it should be noted that we cannot talk about it. Because the resonant islands are not clearly separated and chaotic layers are thick. Let us mention that, unlike the spin-orbit resonance model, the semi-major axis and the eccentricity are not fixed in the spin-orbit coupling model. Thereby, it makes sense to expect different portraits, particularly when they considerably alter, since the widths of the resonances depend on them. Another consequence of altering the semi-major axis and the eccentricity can be seen in Fig. 5, where a chaotic layer is evident at the top of the left panel, while at the top of the right panel the 2:1 resonant island is observable.

For surfing in the phase space of the spin-orbit coupling model, we consider a binary asteroid with values of $a = 10$, $e = 0.05$, $\theta = 0$, $m_s/m_p = 0.05$, and $(a_s : b_s : c_s) = (1 : 0.95 : 0.85)$. We generate cartographic images of the phase space for four categories with different values of orbital semi-major axis, eccentricity, mass ratio, and the semi-axes of the secondary. One parameter is individually changed in each category, and the others are held constant.

### 5.1 Semi-major axis

Let us start with the semi-major axis. Fig. 6 shows the phase portrait for 5, 10, 15, and 20 as the semi-major axis. All FLI maps are almost identical, thus showing that this parameter has no significant effect on the pattern of the phase portrait. The 1:1 and 2:3 resonances are visible in all panels as the main resonance. However, the 1:1 resonance has a considerable resonant island compared with the 2:3 resonance. The others are clear in the bottom panels. Synchronous resonance is abundant in the solar system, particularly since the orbital characteristics of most solar system bodies have a near-circular orbit, regardless of the value of their semi-major axis. The persistence of the 1:3 secondary resonance around the synchronous island center for any value of the semi-major axis is the most interesting one (Wisdom (2004)). By increasing the semi-major axis, the chaotic separatrix of the 1:3 secondary resonance becomes more visible.

In all panels, the interaction between various resonances has led to the emergence of chaos in the phase space. However, we could identify regular and resonant regions in the phase space in addition to chaotic regions.

We select two initial conditions, symbol 'x' in Fig. 6, one in the chaotic zone and the other in the synchronous resonance region to provide time history of normalized angular velocity, $\dot{\phi}/n$, and the synchronous resonance angle, $\psi_1$,respectively, as shown in Fig. 7. The magenta color line in the left panel of Fig. 7 shows the constantly stochastic jumping between different normalized angular velocity over the time. This noise-like time history curve of normalized angular velocity is the characteristic of a chaotic rotation. According to the boundaries of chaotic region in Fig. 6, the value of chaotic normalized angular velocity is almost between 0.5 and 1.75. The upper and lower bounds of time history curve in the left panel Fig. 7 confirm this. The right panel of Fig. 7 illustrates time history of the synchronous resonance angle, $\psi_1$. The oscillatory behavior of the time history curve is due to the eccentricity. The maximum amplitude of the libration is about 10 degrees around the zero equilibrium point.

### 5.2 Eccentricity

Eccentricity is an interesting parameter in analyzing the phase space of the spin-orbit problem. By varying eccentricity, marked differences are visible, and we can see diverse phase space structures. When the eccentricity grows, some resonances emerge, and some resonances vanish. Fig. 8 presents FLI maps of the spin-orbit coupling model for 0.001, 0.01, 0.05, 0.1, 0.2, and 0.5 as the eccentricity. It is evident that as eccentricity increases, the simplicity of the FLI map disappears and the maximum value of FLI increases. This means that the sensitivity to initial conditions is increased for higher eccentricity, and nearby orbits diverge exponentially faster that leads to stronger chaos.

As we can see in the left top panel, for small eccentricity such as $e = 0.001$, the synchronous resonance is dominant in the phase space. The phase space has a pendulum-like structure which confirms the result of the analytical study. However, the thin separatrix layer is chaotic and surrounds the synchronous resonance. This portion of chaos is caused by the homoclinic intersection of stable and unstable manifolds (Khan et al. (1998)).

As shown in the right top panel, the chaotic layer around synchronous resonance is still narrow for $e = 0.01$ yet thicker than that of $e = 0.001$. This is due to overlapping the synchronous resonance from up and 5:4 resonance from down. For this value of eccentricity, the 3:2 resonant region is visible but slight because the size of this resonance depends on the eccentricity, which is small. For $e = 0.01$, the 1:3 secondary resonance appears in the right top panel, expands in the left middle panel, and begins to disappear in the right middle panel. It persists still in the left bottom panel, but chaos finally wipes out these secondary resonance islands in the right bottom panel.

The left middle panel presents the result obtained for $e = 0.05$. For $e = 0.05$, the resonant islands are not separated, i.e., two separatrices of the pendulum-like neighboring resonant regions touch, and the resonances overlap, leading to open diffusion channels between them. Therefore, motion can jump from one resonance to another through the heteroclinic intersection of stable and unstable manifolds of neighborhood resonances.

As Fig. 8 shows, by increasing the eccentricity, the size of the synchronous resonance region shrinks, and the chaotic separatrix layer becomes thick. However, for $e = 0.5$, the size of the synchronous resonance region increases. Also, by increasing eccentricity, high order resonances such as 1:2 resonance begin to appear (see right middle panel). In the last right bottom panel, for $e = 0.5$, the 5:2 resonance also emerges. As explained above, complete overlapping of all resonances naturally produces widespread chaos through Chirikov diffusion (Jafari Nadoushan & Assadian (2016)). However, the resonance islands encircled by a chaotic ocean and the 1:2 resonance island in the bottom of the figure, persist.

We use two initial conditions, the black 'x' symbol in the chaotic zone and the red 'x' symbol in the synchronous resonance region, as depicted in Fig. 8, to provide time history of normalized angular velocity, $\dot{\phi}/n$, and the synchronous resonance angle, $\psi_1$, respectively. The initial condition denoted by black 'x' symbol is surrounded by a vast chaotic ocean. Therefore, we can expect a high-amplitude jump in the time history of normalized angular velocity, as can be seen.

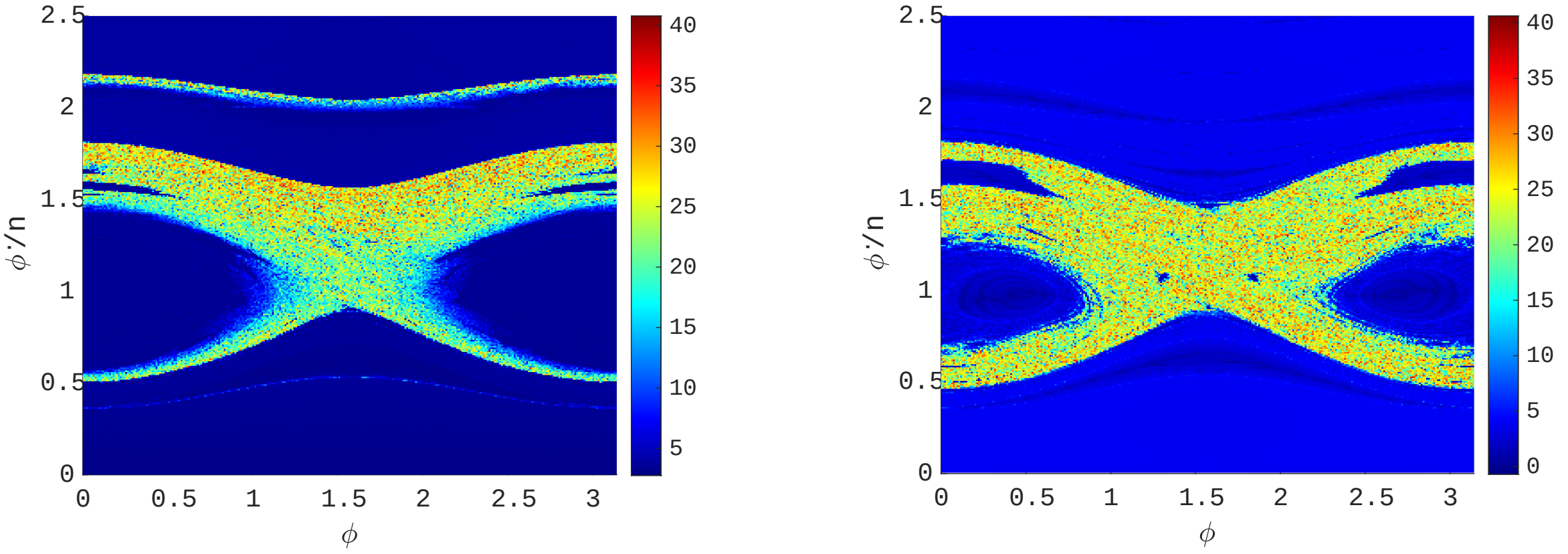

**Figure 5.** FLI maps for spin-orbit coupling (left) and spin-orbit resonance (right).

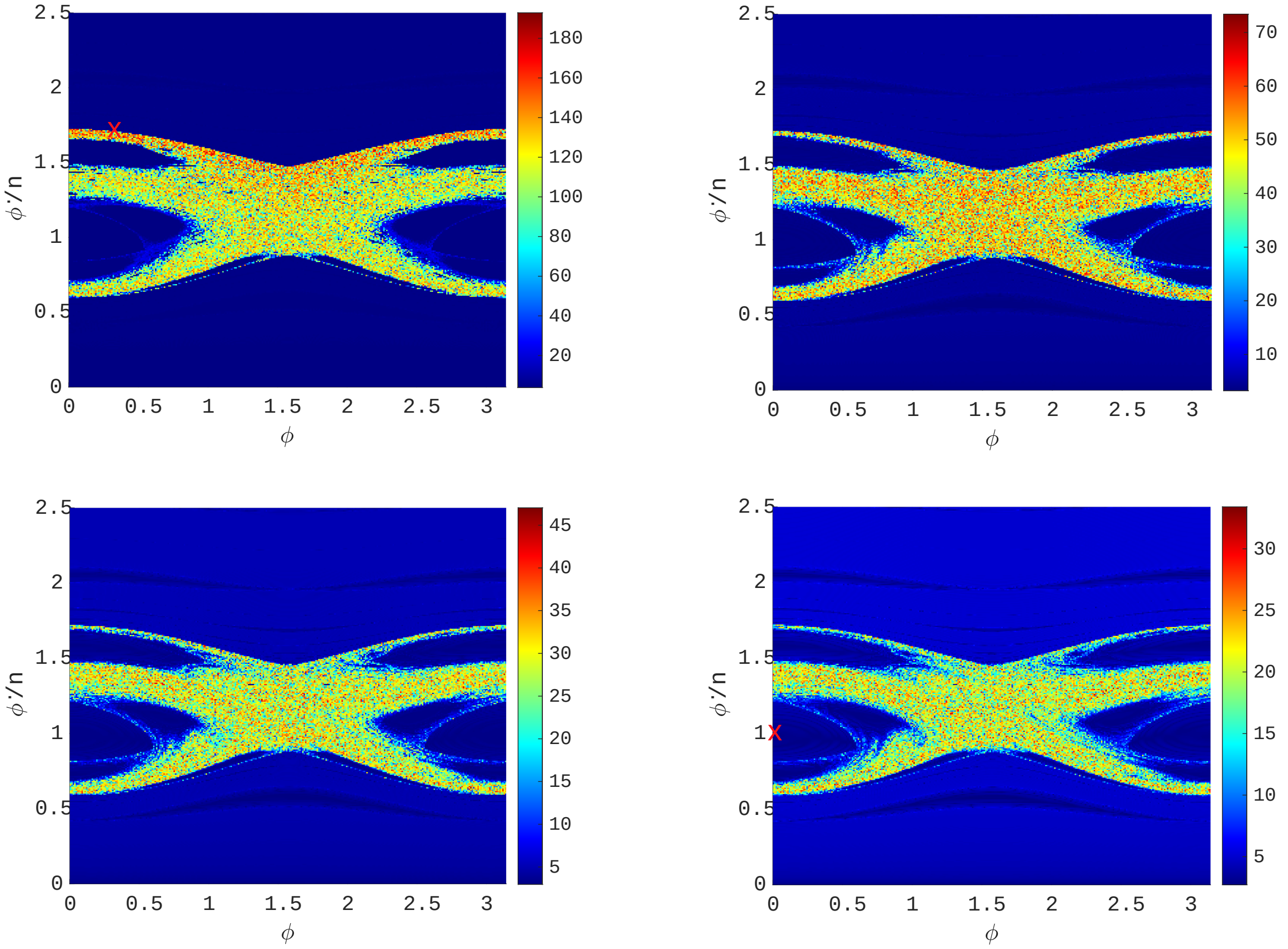

**Figure 6.** FLI maps for different values of semi-major axis 5 (left top panel), 10 (right top panel), 15 (left bottom panel), 20 (right bottom panel). In all panels, the 1:1 and 3:2 resonances surrounding by chaotic ocean, are evident. Also, The 1:3 secondary resonance around the 1:1 resonance is visible. The symbol x in the map, left top and right bottom panels, shows the initial conditions for the trajectories analysed in Fig. 7

However, the time history curve of synchronous resonance angle has a very low amplitude due to the small eccentriciy.

### 5.3 Mass ratio

In this subsection, we study the effect of mass ratio on the phase portrait. Fig. 10 depicts the phase portrait for 0.01, 0.1, 0.5, and 1 as the mass ratio. By increasing the mass ratio, the maximum value of FLI decreases. This means that nearby orbits diverge exponentially slower and, results in weaker chaos. Also the FLI map becomes regular, and the resonant regions become clearly separated. As a result, we are faced with a more regular system in an equal-mass binary system.

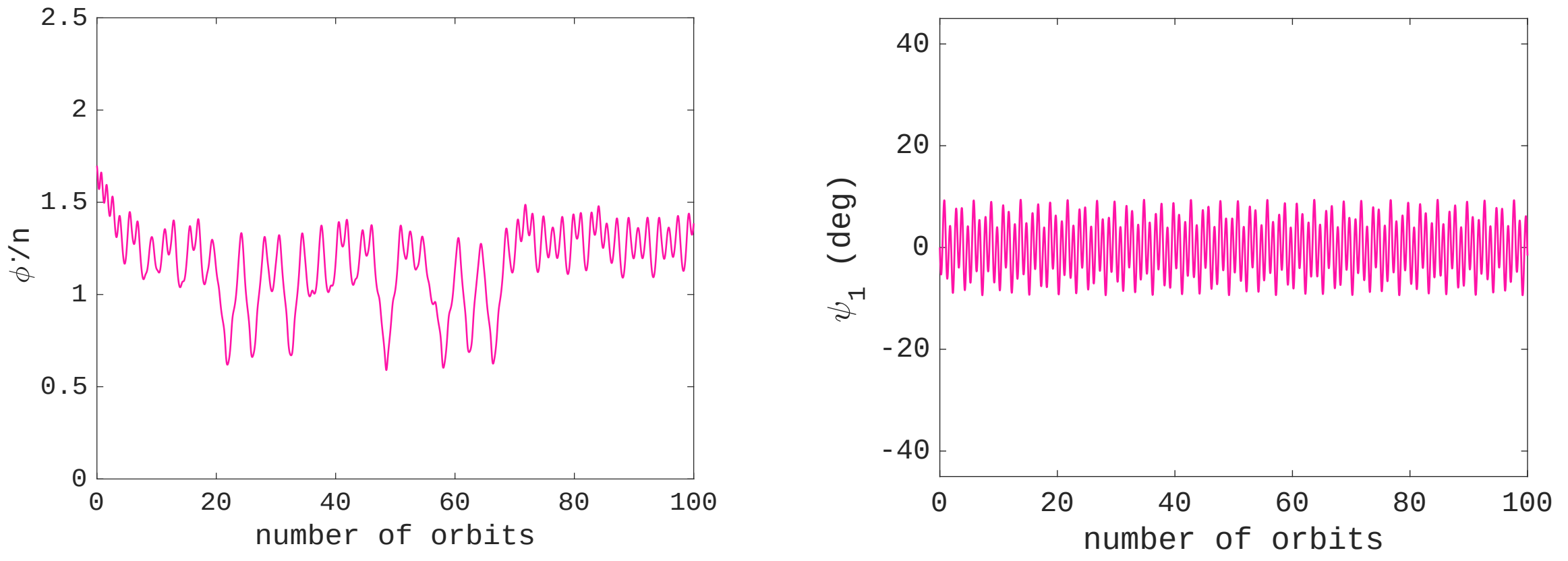

**Figure 7.** The normalized angular velocity (left panel) and the synchronous resonance angle (right panel) versus time for 100 orbits. The initial conditions for normalized angular velocity and resonance angle are shown in the left top and right bottom panels of Fig. 6, respectively, by red x symbol.

As the mass ratio grows, the chaotic separatrix layer of the 3:2 resonance becomes thinner and may disappear. In the left bottom panel, the boundary of 3:2 resonance island is nearly tangent to the boundary of 5:4 resonances. Thus, there is no overlap of resonances. It seems that the synchronous resonance is a major resonance even for eccentric orbits for an equal-mass binary system. Note that the separatrix layer of the synchronous resonance is still chaotic.

The synchronous region contains chains of secondary islands of different periods for different mass ratio values as the bifurcation parameter. As Fig. 10 shows, for the mass ratio of 0.01 and 0.1, we can see the 1:3 secondary resonance chain. By increasing the mass ratio to 0.5, the 1:3 secondary resonance chain disappears, and the 1:4 secondary resonance chain appears. For the mass ratio of 1, there is a 1:5 secondary resonance in the right bottom panel.

The time history curve of normalized angular velocity, $\dot{\phi}/n$, in the left panel of Fig. 11 is generated using the initial condition in the chaotic zone of the left top panel of Fig. 10, which is represented by the black 'x' symbol. This time history curve shows a chaotic behavior, as we expected. The time history curve of synchronous resonance angle, $\psi_1$, in the right panel of Fig. 11 is generated using the initial condition in the synchronous resonance region of the left bottom panel of Fig. 10, which is represented by the red 'x' symbol.

### 5.4 Shape of the secondary

The shape or asphericity of the secondary is the last parameter that we investigate here. It is necessary to emphasize that due to the variable orbit, the $C_{20}$ and $C_{40}$ coefficients are involved in the spin-orbit coupling model, while they are not present in the spin-orbit resonance model. In Fig. 12, we examine the phase portrait considering 1:0.95:0.9, 1:0.9:0.45, 1:0.8:0.8, 1:0.8:0.5, 1:0.6:0.45, 1:0.5:0.25 as the semi-axis ratios. By increasing the asphericity, the maximum value of FLI increases and the system experiences stronger chaos. Also, the more the shape of the secondary deviates from a sphere, the more chaotic ocean grows, and simultaneously all resonances, except the synchronous one, begin to deform considerably and then disappear.

Notice that the asphericity parameters exist at coefficients of all resonance, particularly the synchronous resonance. Therefore, the asphericity has a significant contribution to the width of the synchronous resonances in such a way that the width of synchronous resonance grows progressively by decreasing the semi-axis ratios. As a consequence of the non-spherical nature of asteroids, synchronous resonance could dominate the phase space and is the main resonance in the binary asteroid systems. As the left and right bottom panels show, when the secondary has a more irregular shape, a large stable structure can be detected inside the liberation region of synchronous resonance that leads to this resonance playing the main role in the dynamics of the system. However, the chaotic area is also developed, and wide spread chaos occurs which leads to more instability.

Fig. 12 suggests that the synchronous resonant region encompasses a chain of 1:3 and 1:2 secondary islands. As the asphericity increases, we see that the 1:3 secondary resonance (left top panel) bifurcates to 1:2 (right top panel) and then bifurcates to 1:3 secondary resonances (left middle panel). Let us examine this in more detail. In Fig. 13, we consider the semi-axis ratio values between the values of the first and the third cases of Fig. 12, i.e., 1:0.925:0.9, 1:0.9:0.9, 1:0.875:0.85, 1:0.85:0.85. It is evident that the 1:3 secondary resonance in the left top panel of Fig. 12 bifurcates to the 1:5 secondary resonance in the left top panel of Fig. 13, then the 1:5 bifurcates to the 1:2. As the left bottom panel depicts, the 1:2 secondary resonance starts being wiped out by the chaotic ocean. Finally, the 1:2 secondary resonance in the right bottom panel of Fig. 13 bifurcates to 1:3 in the left middle panel of Fig. 12. These bifurcations can be the result of material transfer from one body to another and changing the shape of the secondary.

At the same time, by increasing asphericity, changes in the primary resonances are apparent. The 3:2 resonance shrinks while the synchronous resonance area grows. The 1:2 resonance finally disappears, as seen in the last panel of Fig. 12.

In Fig. 12, the initial condition denoted by black 'x' symbol is at the heart of the chaotic ocean. Thus, we see the high-amplitude and relatively high-frequency constantly stochastic jumping in the time history of normalized angular velocity (the left panel of Fig. 14) that indicates the rotation of the secondary is more chaotic. We use the initial condition, the red 'x' symbol in the synchronous resonance

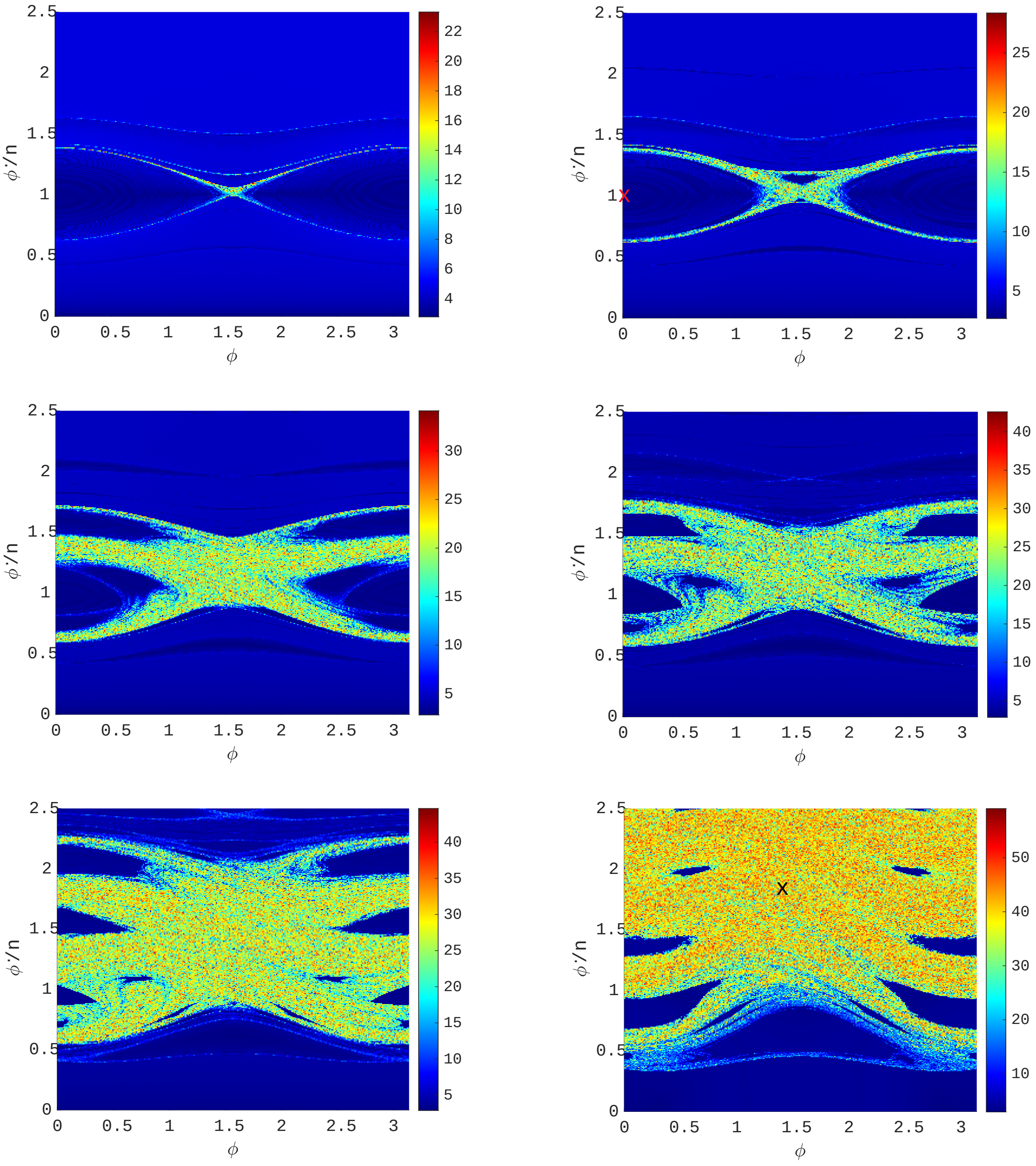

**Figure 8.** FLI maps for different values of eccentricity 0.001 (left top panel), 0.01 (right top panel), 0.05 (left middle panel), 0.1 (right middle panel), 0.2 (left bottom panel), 0.5 (right bottom panel). The 1:1 resonance is visible in the top panels. In the middle panels, in addition to the main resonances, the 1:3 secondary resonance is also observed. The 2:1 resonance is visible in the bottom panels. The symbol x in the map, right top and right bottom panels, shows the initial conditions for the trajectories analysed in Fig. 9

region, to provide the time history of the synchronous resonance angle, $\psi_1$, as depicted in the right panel of Fig. 14. An increase in asphericity seems to increase the amplitude of the libration which its maximum is about 20 degrees around the zero equilibrium point.

## 6 (65803) DIDYMOS BINARY ASTEROID

In this section, we study the binary near-Earth asteroid (65803) Didymos, including Didymos as the primary and Dimorphos as the secondary, as an example. The (65803) Didymos is the target of the first planetary defense mission known as DART (Cheng et al. (2012)).

The orbital and physical parameters of this binary asteroid are $a = 12.6316$, $e = 0.025$, $m_s/m_p = 0.00935$, and $(a_s : b_s : c_s) = (1 : 0.8334 : 0.7576)$ (Agrusa et al. (2022)). It should be noted that the parameters are non-dimentional. According to the previous section, the relatively large semi-major axis makes the phase space tend towards order. However, eccentric orbit, small mass ratio, and large asphericity make the phase space chaotic. With regards to the orbital and physical parameters of the system and criterion developed in equation 28, the resonances overlap, and chaos is inevitable. Although, due to overlapping resonances, the hypothesis of isolated resonances is not true here, it can provide some insight into the phase space.

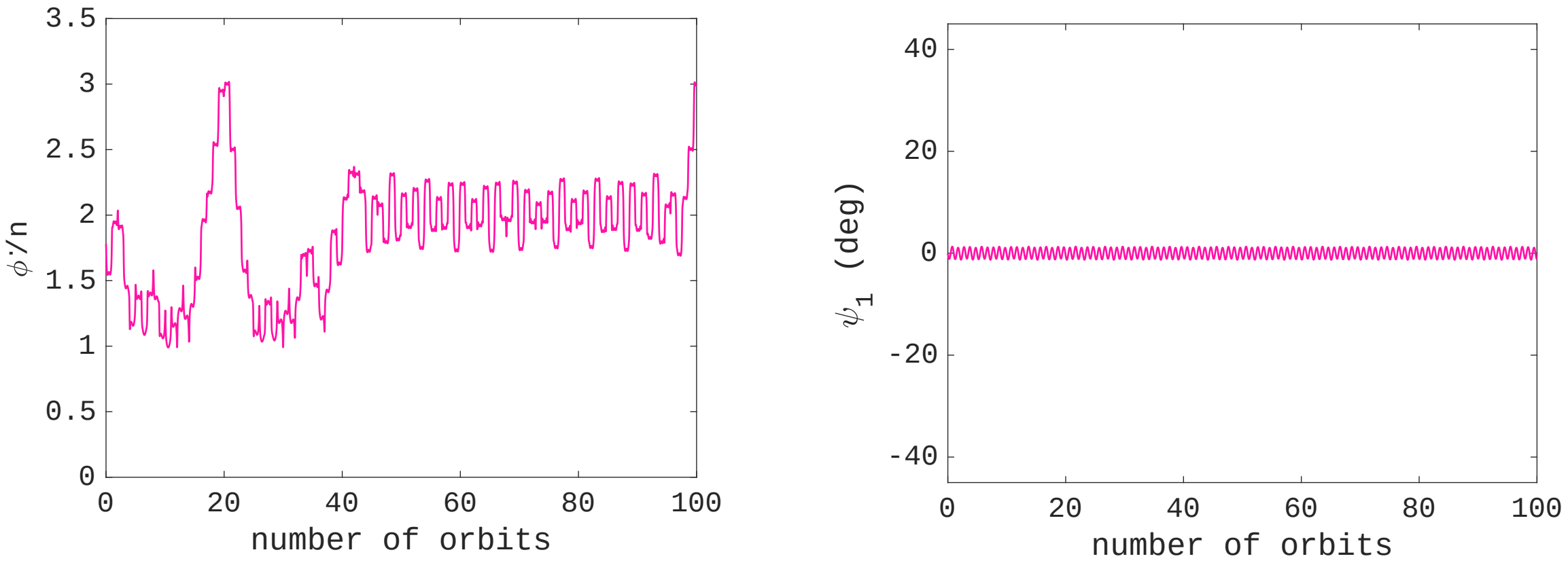

**Figure 9.** The normalized angular velocity (left panel) and the synchronous resonance angle (right panel) versus time for 100 orbits. The initial conditions for normalized angular velocity and resonance angle are respectively shown by black and red x symbol in Fig. 8.

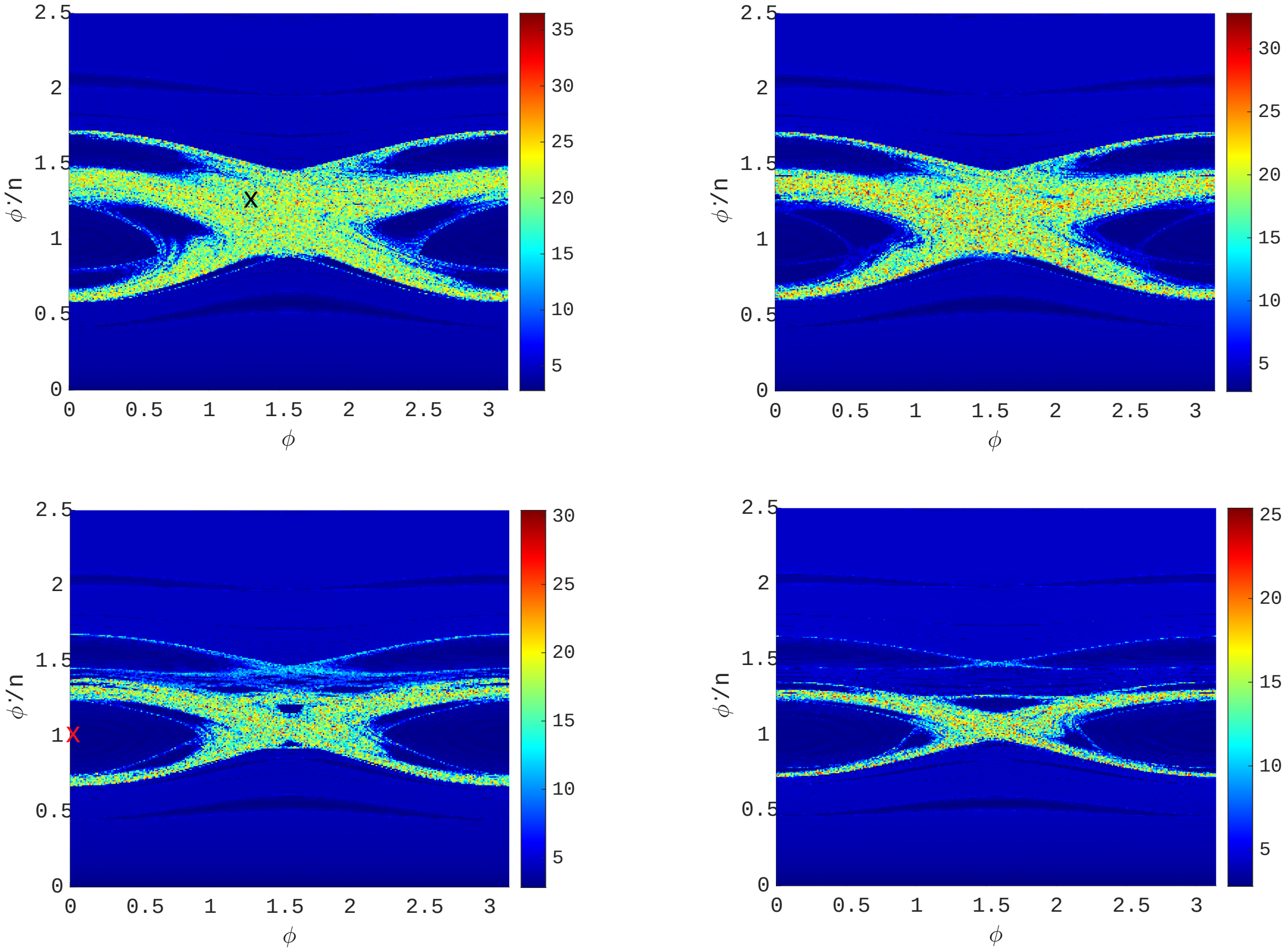

**Figure 10.** FLI maps for different values of mass ratio 0.01 (left top panel), 0.1 (right top panel), 0.5 (left bottom panel), 1 (right bottom panel). In all panels, the 1:1 and 3:2 resonances are visible. The 1:3 secondary resonance in the top panels bifurcates to the 1:4 secondary resonance in the left bottom panel and to the 1:5 secondary resonance in the right bottom panel. The symbol x in the map, left top and left bottom panels, shows the initial conditions for the trajectories analysed in Fig. 11

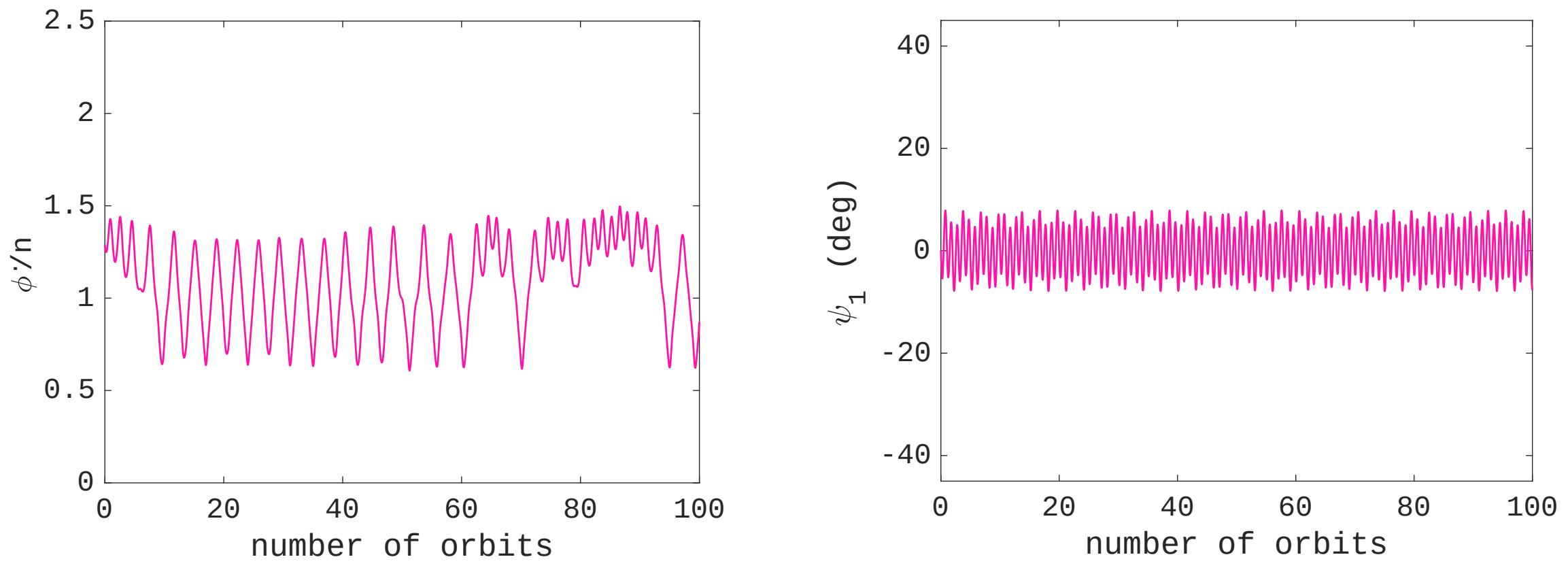

**Figure 11.** The normalized angular velocity (left panel) and the synchronous resonance angle (right panel) versus time for 100 orbits. The initial conditions for normalized angular velocity and resonance angle are respectively shown by black and red x symbol in Fig. 10.

Fig.15 shows the reduced phase space for the (65803) Didymos binary asteroid. The resonant islands of the 2:1, 1:1, and 2:3 resonances are distinctly shown in the figure using analytical method. As said before, the asphericity has a major effect on the width of all resonances. Hence, the 1:1 resonant island is too wide and has spread almost over the phase space. The width of the other resonances are scaled by eccentricity. Therefore, the 2:1 and 2:3 resonances are less wide than the 1:1 resonance, but they overlap, as illustrated in Fig.15. Given the interaction of the resonances, we expect to see chaos in the phase space. Furthermore, since the width of the resonance indicates the strength of the resonance, we can conclude that the 1:1 resonance plays the main role in the phase space and hereupon, in the dynamics of the system.

To have a correct image of the phase space and thus dynamics of the system, we provide the FLI map in Fig.16. It is evident that due to the strength of 1:1 resonance, i.e., the relatively sizeable resonant region, the (65803) Didymos binary asteroid could be in synchronous state as it is (Pravec et al. (2019)). The eccentric orbit and relatively large asphericity of the secondary can cause its rotation to become chaotic, as seen in Fig.16.The chaotic region in the phase space is caused by the overlap of the resonances. However, there are some stable islands encircled by the chaotic ocean, including the displaced narrow zone of 2:3 resonance.

## 7 CONCLUSIONS

Based on the definition of close binary asteroid systems, that reflects dynamical closeness of asteroid pair, when the shape of the secondary is irregular, coupling of orbital and rotational motions should be taken into account to capture the correct dynamics of a binary system. In this paper, we presented analytical and numerical studies of phase space in the spin-orbit coupling model through geographic and cartographic information. To this end, we provided the Hamiltonian of the system, including gravitational potential function of a binary system composed of a spherical primary and an ellipsoidal secondary. Geographic information of the resonances says that the orbital semi-major axis has a more prominent role because of variable orbit. Therefore, the amplitude of resonances was modified in the spin-orbit coupling model. Thus, the overlap of resonances might happen with moderate orbital elements. The analytical result, confirmed by the numerical result, showed that the synchronous resonant island is wider when dealing with the spin-orbit coupling model. However, the rotational dynamics of the secondary is complicated, and it is influenced by various system parameters. As the numerical study showed, by increasing the mass ratio, the maximum value of FLI decreases. On the other hand, by increasing eccentricity and asphericity, the maximum value of FLI increases. The orbital semi-major axis have no significant effect on the pattern of the phase portrait. Nevertheless, by changing eccentricity, diverse phase space structures appeared. When the mass ratio increased, resonant regions became clearly separated, and we saw a more regular system in an equal-mass binary system. When the asphericity of the secondary increased, the chaotic ocean grew, and most resonances began to deform substantially. Since the asphericity parameters exist at coefficients of all resonances, especially the synchronous one, the width of synchronous resonance grew progressively by decreasing the semi-axis ratios. As a result of irregular shape of asteroids, synchronous resonance is prevalent in binary asteroid systems. Furthermore, we observed some bifurcations in the secondary resonance when the semi-axis ratios decreased, which can be caused by material transfer. Finally, the (65803) Didymos binary asteroid was taken as a case to show analytical and numerical results.

## ACKNOWLEDGEMENTS

The author thanks the anonymous reviewer whose valuable suggestions and comments helped improving this paper.

## DATA AVAILABILITY

The data underlying this paper are available from the corresponding author on reasonable request.

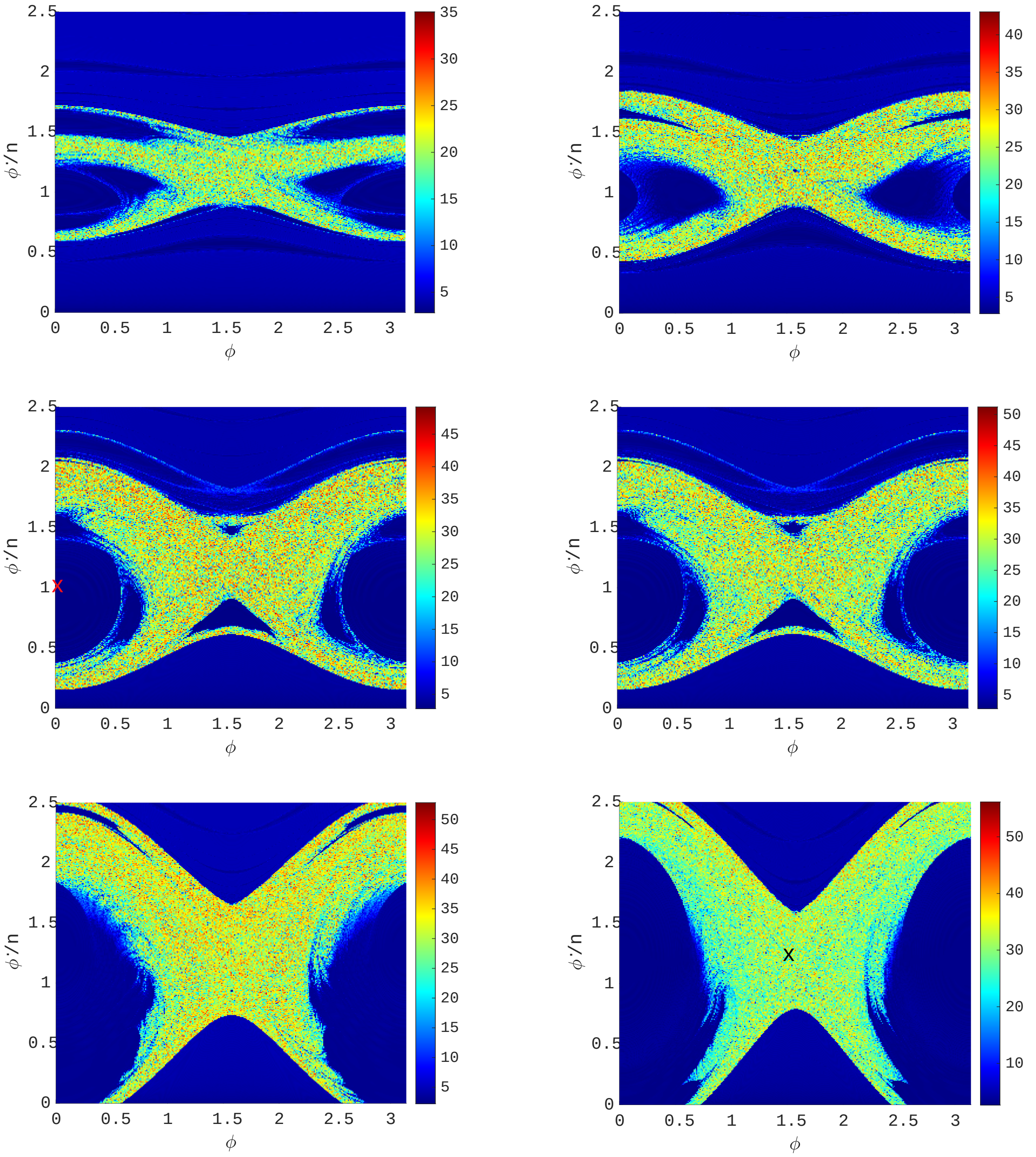

**Figure 12.** FLI maps for different values of the semi-axes of the secondary 1:0.95:0.9 (left top panel), 1:0.9:0.45 (right top panel), 1:0.8:0.8 (left middle panel), 1:0.8:0.5 (right middle panel), 1:0.6:0.45 (left bottom panel), 1:0.5:0.25 (right bottom panel). The 1:3 secondary resonance in the left top panel bifurcates to the 1:2 secondary resonance in the right top panel and again bifurcates to the 1:3 secondary resonance in the left middle panel. This secondary resonance disappear in the bottom panels. The symbol x in the map, left middle and right bottom panels, shows the initial conditions for the trajectories analysed in Fig. 14

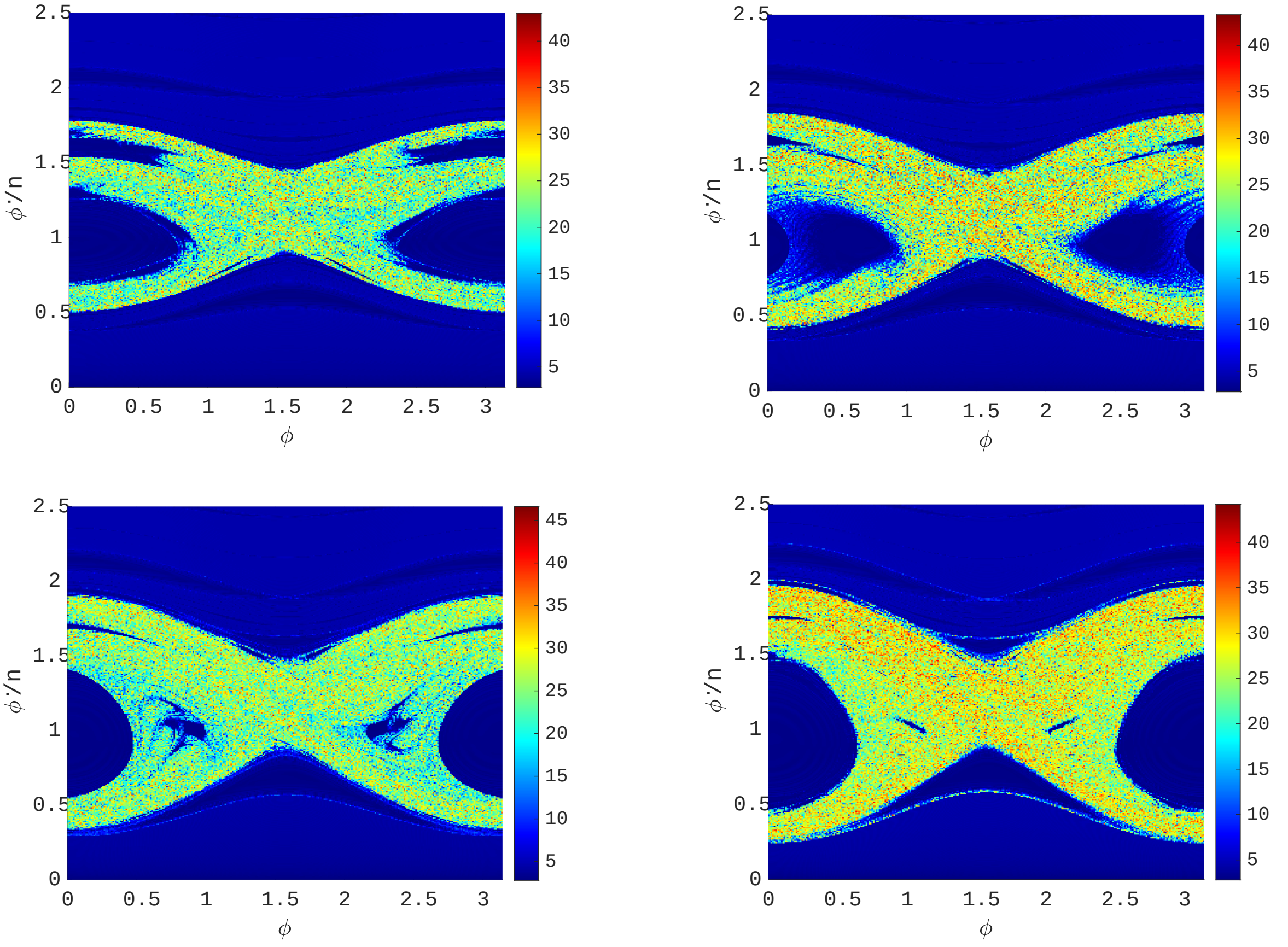

**Figure 13.** FLI maps for different values of the semi-axes of the secondary 1:0.925:0.9 (left top panel), 1:0.9:0.9 (right top panel), 1:0.875:0.85 (left bottom panel), 1:0.85:0.85 (right bottom panel).

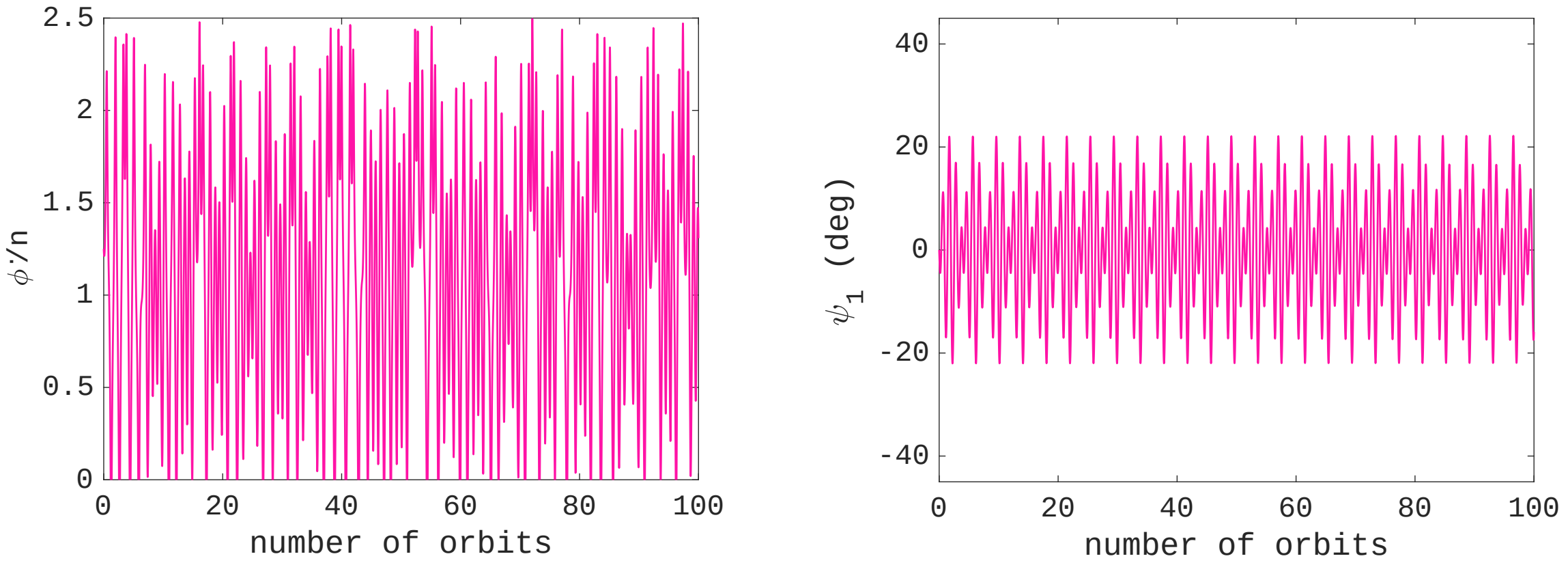

**Figure 14.** The normalized angular velocity (left panel) and the synchronous resonance angle (right panel) versus time for 100 orbits. The initial conditions for normalized angular velocity and resonance angle are respectively shown by black and red x symbol in Fig. 12.

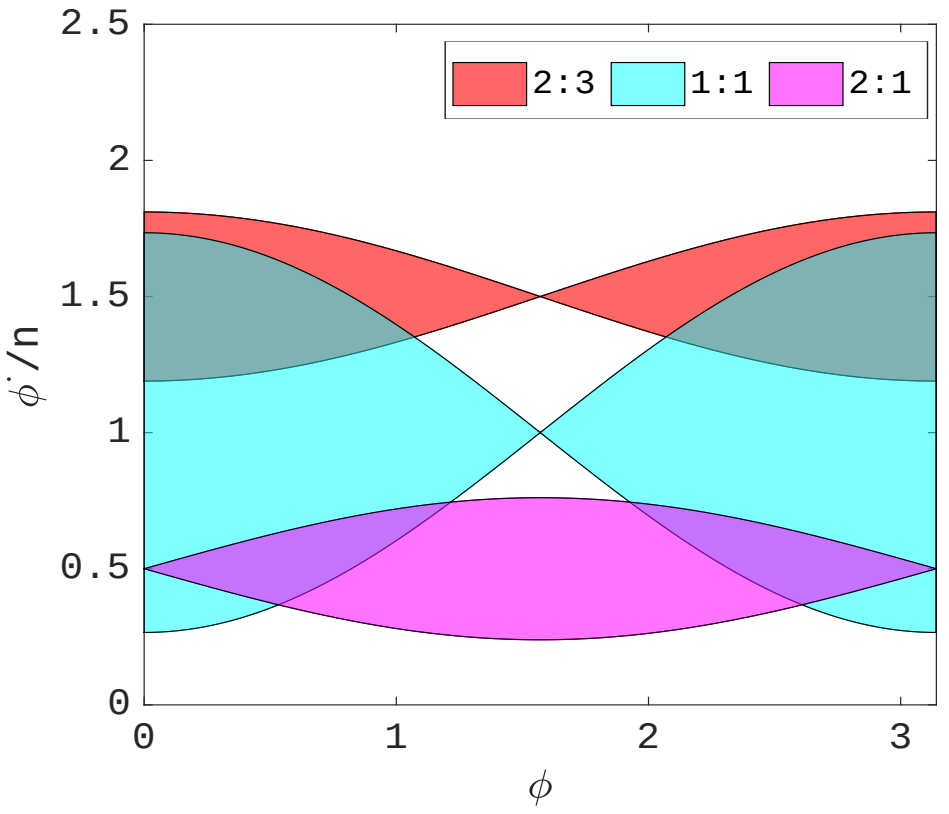

**Figure 15.** Reduced phase space of (65803) Didymos binary asteroid.

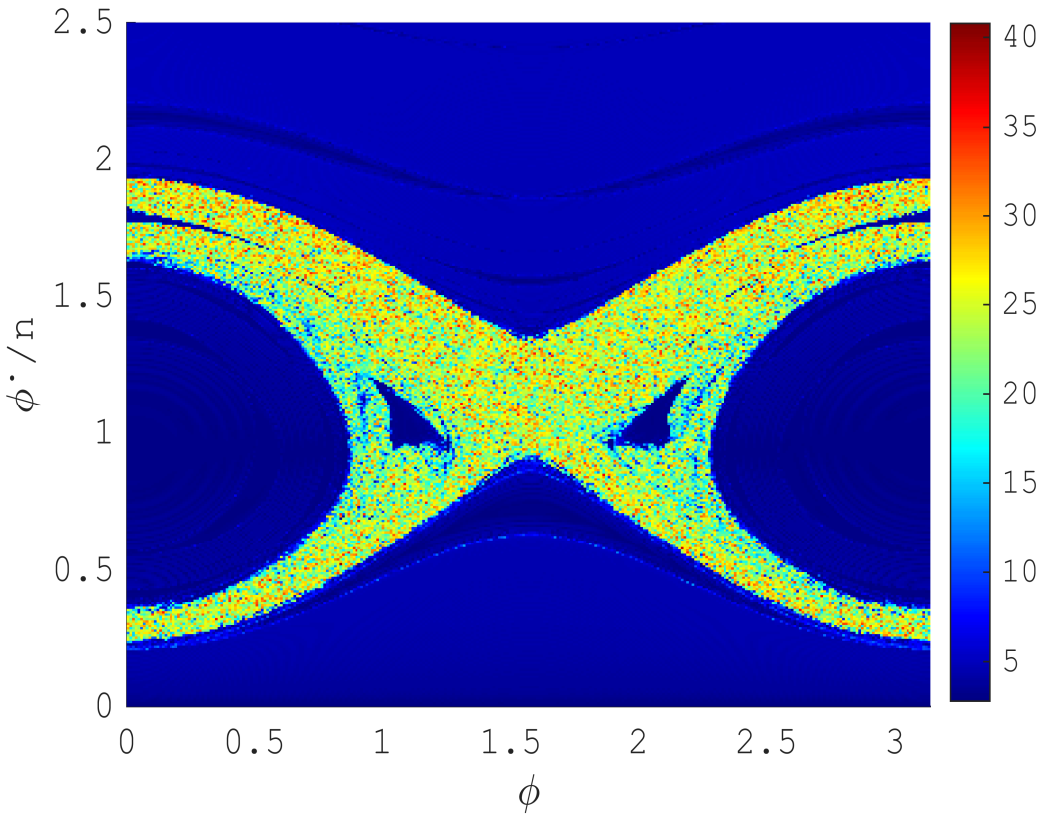

**Figure 16.** FLI maps of (65803) Didymos binary asteroid.

This paper has been typeset from a TEX/LATEX file prepared by the author.